\documentclass[aps,prd,twocolumn,,preprintnumbers,groupedaddress,longbibliography,showkeys,showpacs]{revtex4-1}

\usepackage{graphicx}
\usepackage{bm}
\usepackage{amssymb,amsfonts}
\usepackage{amsmath}
\usepackage{amsthm}
\usepackage{graphicx}
\usepackage[usenames,dvipsnames]{xcolor}
\usepackage{soul}
\usepackage{url}
\graphicspath{{Figures/}}

\newcommand{\one}{1 \hspace{-1.2mm}  {1}}

\begin{document}

\title{Hidden symmetries, spin and charge of artificial magnetic monopoles}

\author{Alexander I. Nesterov}
   \email{nesterov@cencar.udg.mx}
\affiliation{Departamento de F{\'\i}sica, CUCEI, Universidad de Guadalajara,
 Guadalajara, CP 44420, Jalisco, M\'exico}
 
\author{Gennady P  Berman}
 \email{gpb@lanl.gov}
\affiliation{Theoretical Division, Los Alamos National Laboratory,  Los Alamos, NM 87545, USA}

\date{\today}

\begin{abstract}
We discuss the non-Abelian artificial magnetic monopoles associated with $n$-level 
energy crossing in quantum systems. We found that hidden symmetries reveal themselves as 
observables such as spin, charge and other physical degrees of freedom. We illustrated our 
results on concrete examples of two and three energy-level  crossing. Our results can be useful 
for modeling of various phenomena in physical and biological systems.
\end{abstract}


\preprint{LA-UR-21-23554}

\maketitle

\section{Introduction}
\label{intro}

Recent experimental results provide evidence for the emergence of artificial magnetic monopoles in the crystal-momentum space, noncommutative quantum mechanics, the anomalous Hall effect of ferromagnetic materials, and magnetic superconductors, trapped $\Lambda$-type atoms, anisotropic spin systems, in a Bose-Einstein condensate of degenerate dressed states, etc.  In particular, these monopoles own artificial magnetic fields, making it possible to model many physical phenomena \cite{BM,Br,FNT,SR,Hal,FP,ZLS,MSN,PVMM,DGJ,RPN,NVJ,RVWC,CSPA,RRKMH}. 

These artificial magnetic monopoles are related to the Berry phase, and energy-level crossing \cite{B0,B1}. The effective gauge field gives rise to a `magnetic' monopole located at the parameter space's degeneracy point. In the simplest case of double degeneracy with two linearly independent eigenvectors, the energy surfaces form a double cone's sheets. In the neighborhood of the degenerate point (apex of the cones), the behavior of the system can be characterized by three parameters: $\mathbf R= (X, Y, Z)$ \cite{Arn}. Each eigenstate, $|n,\mathbf R \rangle$, gives rise to the Berry's connection defined by ${\mathbf A}_n(\mathbf R)= i\langle n,\mathbf R| \nabla_{\mathbf R} |n,\mathbf R \rangle$, and the curvature $\mathbf B_n = \nabla_{\mathbf R} \times {\mathbf A}_n $ associated with ${\mathbf A}_n$, is the field strength of artificial magnetic monopole located at the degenerate point \cite{B0,B1,BW,BD}. 

In general, for a quantum-mechanical system governed by a slowly varying parameter-dependent and periodic Hamiltonian, the wavefunction gains a geometric phase, which depends on the path in parameter space. For a non-adiabatic evolution of the quantum system,  Berry's phase becomes the Aharonov-Anandan geometric phase \cite{AA}.  The further development of this concept for the quantum mechanical systems with degenerate eigenvalues leads to the non-Abelian geometric phases \cite{WFZA,MCA,MCA1}. (For a comprehensive review, see \cite{WFSA}.) With each eigenstate, a monopole is associated, which is located at the degenerate point and carries its own charge. However, the total charge equals zero. For $n$-level energy crossing, one has $n$ monopoles with different charges. And again, the total charge equals zero. 

In this paper, we show that instead of $n$-monopoles with the different charges, $n$-level energy crossing can be characterized by a single non-Abelian artificial monopole 
located at the degenerate point. The monopole has matrix-valued ``charge" with values in the Lie algebra of the group $SU(n)$. In particular, for trapped $\Lambda$- type atom interacting with two laser beams, the $SU(3)$ monopole emerges. We show that in this case, the hidden symmetry reveals itself not as the spin of the monopole but as a rather complicated mixture of spin and charge degrees of freedom.

The structure of the paper is as follows. In Sec. II, we describe the Dirac magnetic monopole's 
general properties and discuss its spin. In Sec. III, we discuss the relationship between artificial 
magnetic monopoles and geometric phases. We show how the isospin of the artificial magnetic 
monopole turns into the monopole's spin. In Sec. IV, we discuss the properties of quasienergies 
and geometric phases for time-periodic quantum systems. In particular, we consider 
superconducting qubit interacting with resonator, and spin system driven by circularly polarized 
field, to illustrate our findings. In Sec V, we generalize our results for $n$-level  energy 
crossing. We demonstrate that in this case, a non-Abelian single fictitious magnetic monopole with a matrix-valued charge occurs. In Conclusion, we discuss our results and possible applications. Through the paper, unless stated otherwise, we use the natural units, 
$\hslash=c=1$.

\section{Mathematical preliminaries}
\label{Math}


A magnetic monopole, introduced by Dirac in 1931 \cite{Dir}, yields a magnetic field of pointlike particle,
\begin{equation}
{\mathbf B} = q \frac{\mathbf r}{r^3},
\label{eq_0}
\end{equation}
where $q$ is a charge of monopole. From the second pair of Maxwell equations it follows,
\begin{align}
\nabla \cdot {\mathbf B} = 4\pi q\delta^3(\mathbf r).
\label{eq_1}
\end{align}
However, the definition $\mathbf B = \nabla \times \mathbf A$ implies that $\nabla \cdot {\mathbf B}
  =0$, that is in a contradiction with Eq. (\ref{eq_1}). This contradiction can be resolved by using a singular vector potential, 
  \begin{equation}
{\mathbf A}_{\mathbf n}= q\frac{{\mathbf r}\times {\mathbf n}}
{r(r - {\mathbf r} \cdot{\mathbf n})}.
\label{d_str}
\end{equation}
Here the unit vector, $\mathbf n$, determines the direction of a singular string passing from the origin of coordinates to infinity. It is easy to verify now that ${\mathbf B} ={\rm rot}{\mathbf A} + {\mathbf h}$,
where
\begin{eqnarray}
{\mathbf h} = 4\pi q{\mathbf n}\int _{0}^\infty
\delta(\mathbf r - \mathbf n \tau) d \tau 
\end{eqnarray}
is the contribution of the singular string.

The choice of the vector potential (\ref{d_str}) is unique up to a gauge transformation. For instance, the Schwinger's choice is  \cite{Sw_1}
\begin{equation}
{\mathbf A^{SW}}= q\frac{({\mathbf n}\cdot {\mathbf r}){\mathbf r}\times {\mathbf n}}
{r(r^2 - ({\mathbf n} \cdot{\mathbf r})^2)} , 
\label{sw}
\end{equation}
and the string is  propagated from $-\infty$ to $\infty$, so that
\begin{eqnarray}
{\mathbf h^{SW}} = 2\pi q{\mathbf n}\int _{-\infty}^\infty
\delta(\mathbf r - \mathbf n \tau) d \tau .
\end{eqnarray}
Using the gauge freedom in the definition of the vector potential, one can show that \cite{NF,N1a}
\begin{align}
{\mathbf A^{SW}} = {\mathbf A}_{\mathbf n} + \nabla \chi_{\mathbf n},
\end{align}
where
\begin{eqnarray}
\nabla\chi_{\mathbf n} = q\frac{\mathbf n\times
\mathbf r}{r^2- (\mathbf n \cdot \mathbf r)^2},
\label{A_02}
\end{eqnarray}
with $\chi_{\mathbf n}$ being polar angle in the plane orthogonal to
${\mathbf n}$.
The Dirac and Schwinger vector-potentials are related as follows:
\begin{align}
{\mathbf A^{SW}} = \frac{1}{2}({\mathbf A}_{\mathbf n} + {\mathbf A}_{-\mathbf n}).
\end{align}

From rotational symmetry of the theory, it follows immediately that an arbitrary gauge transformation ${\mathbf A}_{\mathbf n} \rightarrow
{\mathbf A}_{\mathbf n'}$ can be undone by rotation $\mathbf r
\rightarrow g \mathbf r$, $g\in\rm SO(3)$. Using this fact and adapting the results of \cite{Wu2,Jac,Jac1}, one can show that the compensating gauge transformation $U_g$ is defined by a nonintegrable phase factor,
\begin{align}
&U_g\Psi(\mathbf r) =
\exp(i\alpha(\mathbf r, \mathbf
n;g))\Psi (\mathbf r),
\label{g_0}\\
&\alpha(\mathbf r, \mathbf n;g)= e \int_{\mathbf
r}^{\mathbf r'} \mathbf A_{\mathbf n}(\boldsymbol \xi)
\cdot d \boldsymbol \xi, \quad \mathbf r' = g \mathbf r,
\label{g_1c}
\end{align}
where the integral is taken along the geodesic
$\widehat{\mathbf  r \,\mathbf r'}\subset S^2$. 

The equation of motion of a non-relativistic particle of a charge $e$ and mass $m$, moving in the field of a magnetic monopole, reads 
\begin{equation}
m\ddot{\mathbf r} = \frac{\mu}{r^3}{\mathbf r} \times\dot{\mathbf r},
\label{Eq1}
\end{equation}
where $\mu = eq$.  From Eq. (\ref{Eq1}) one can deduce that 
\begin{align}
\frac{d}{dt } {\mathbf J} = 0, \quad \frac{d}{dt } {\mathbf J}\cdot {\mathbf S} = 0 \quad \text{and} \quad\frac{d}{dt } {\mathbf J}\cdot {\mathbf L}_g = 0.
\end{align}
Here ${\mathbf L}^g ={\mathbf r} \times \left({\mathbf p} - e{\mathbf A}\right)$ is a gauge invariant angular momentum, and
\begin{align}
{\mathbf J} = {\mathbf L}^g+ {\mathbf S},
\label{EQ}
\end{align}
is a  total angular momentum, ${\mathbf S} = -\mu \hat{\mathbf r} $ with $\hat{\mathbf r} $ being a unit vector from $q$ to $e$.

\begin{figure}[tbh]
\scalebox{0.575}{\includegraphics{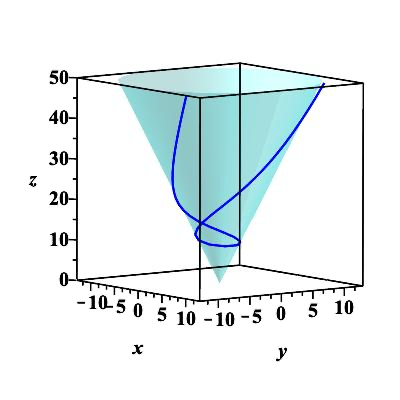}}
\caption{The trajectory of a charged particle in the field of Dirac monopole.
\label{DP1a}}
\end{figure}

One can show, by making use of Eqs. (\ref{Eq1}), (\ref{EQ}), that 
the motion of the charged particle is confined to a cone with axis $\mathbf J$ and angle $2\theta_0$ (see Fig. \ref{DP1a}) \cite{NG,LRP,LBNZ,SHY}
\begin{align}
\cos\theta_0 = \frac{\mu}{J}.
\end{align}
The unit vector $\hat{\mathbf r}$ precesses around $\mathbf J$ with the angular velocity 
\begin{align}
\omega = \frac{J}{mr^2}.
\end{align}
The motion of the charged particle in the field of the magnetic monopole is unbounded (Fig. \ref{DP1a}), and taking into account the conservation of the energy, one can write \cite{SHY}
\begin{align}
r=\sqrt{ v^2t^2 + b^2},
\end{align}
where $b$ is the minimal distance between the particle and monopole located at the origin of coordinates. The equation of orbit can be written as,
\begin{align}
\frac{b}{r} = \cos\bigg(  \frac{L^g}{J}\varphi\bigg), \quad -(J/L^g)\pi/2 < \varphi < (J/L^g)\pi/2 ,
\end{align}
where $L^g = |{\mathbf L}^g|$.

 In the framework of the quantum mechanics, a motion of charged massive non-relativistic particle in the field of a magnetic monopole is described by the Schr\"odinger equation 
\begin{align}
i\frac{\partial}{\partial t}|\Psi \rangle= H |\Psi \rangle ,
\end{align}
where the Hamiltonian is given by
\begin{align}
H = \frac{1}{2m} ({\mathbf p} - e{\mathbf A})^2.
\end{align}
 Dirac showed that quantum mechanics is consistent with the existence of magnetic monopoles, if  the quantization condition $2\mu  =n$, with $n$ being integer, holds \cite{Dir}. 

Writing the wavefunction as $|\Psi \rangle = e^{-iEt}|\psi \rangle$ and employing the spherical coordinates, we obtain for eigenvalue problem, $H|\psi \rangle = E|\psi \rangle$, the following equation,
\begin{equation} 
\bigg(-\frac{1}{2mr^2}\frac{\partial}{\partial r} \bigg( r^2
\frac{\partial}{\partial r}\bigg)+ \frac{({\mathbf J}^2 -
\mu^2)}{2mr^2} \bigg)| \psi \rangle = E |\psi \rangle.
\label{eq01}
\end{equation}

In a classical approach the last term in Eq.(\ref{EQ}) usually is interpreted as a contribution of the electromagnetic field, which
carries an angular momentum \cite{WHA,Gol1,Gol2},
\begin{align}\label{L1a}
{\mathbf L}_{em}=\frac{1}{8\pi}\int {\mathbf r'} \times({\mathbf E}\times{\mathbf B}) d^3 r' = 
- \mu\hat{\mathbf r} .
\end{align}
A quantum-mechanical description leads to the generalization of this picture and interpretation of 
${\mathbf S} $ as a spin of the composed charge-monopole particle\footnote{The notion of an intrinsic spin in the Dirac monopole problem  was introduced by Goldberg \cite{Gol1,Gol2}. The role of spin in the consistent theory of magnetic monopole is widely discussed in the literature, see, e.g. \cite{WHA,Gol1,Gol2,SMN,SMN1,WF,WF1,JRRC,HPHG,SJ1,MKD,MAK}.}. 

Classical total angular momentum (\ref{EQ}) can be recast as
\begin{align}
{\mathbf J} = m{\mathbf r}\times {\mathbf v} - \mu \hat{\mathbf r}.
\end{align}
Motivated by this classical expression, one could introduce the spin by defining the momentum operator $\mathbf p$ as \cite{SJ1},
\begin{align}
m{\mathbf v} = {\mathbf p} + {\mathbf S }\times {\hat{\mathbf r}}/r,
\end{align}
which yields
\begin{align}
{\mathbf J} = {\mathbf r}\times {\mathbf p} + {\mathbf S}={\mathbf L} + {\mathbf S}.
\label{J1}
\end{align}
Here ${\mathbf L} = -i{\mathbf r} \times \nabla$ is a standard angular momentum operator, and $\mathbf S$ obeyes usual spin commutation relations:
\begin{align}
[S_i, S_j] =i \varepsilon_{ijk} S_k,\, \quad [L_i, S_j] =0.
\end{align}
Then one can see that the gauge invariant operator ${\mathbf L}^g$ leads to the following commutation relations:
\begin{align}
[L^g_i, L^g_j] =i \varepsilon_{ijk} (L^g_k +S_k),
\end{align}
and, thus, it is not acceptable as the angular momentum for a charged particle moving in the field of a magnetic monopole. The correct commutation relations are obtained for the total angular momentum:
\begin{eqnarray}
[J_i, J_j] = i\epsilon_{ijk}J_k, \quad [{\mathbf J}^2, J_i] =
0.
 \label{eq5}
\end{eqnarray}

Computation of the kinetic energy, $T=m {\mathbf v}^2/2$, yields \cite{MAK}
\begin{align}
H = \frac{1}{2m}p^2_r + \frac{1}{2m r^2}({\mathbf J}^2 - ({\mathbf S}\cdot {\hat{\mathbf r}})^2 ),
\label{EQ1}
\end{align}
where
\begin{align}
p^2_r = \frac{1}{r^2} \big (  (\mathbf r \cdot \mathbf p)^2   -i \mathbf r \cdot \mathbf p \big ).
\label{EQ1a}
\end{align}
From (\ref{EQ1}) it follows
\begin{align}
[H,{\mathbf S}\cdot\hat{\mathbf r}] =0.
\label{EQ2}
\end{align}
 Using (\ref{EQ2}), one can look for a solution of Eq. (\ref{eq01}) obeying
\begin{align} \label{S1g}
& ({\mathbf S}\cdot\hat{\mathbf r})\,|\psi \rangle = -\mu |\psi \rangle,  \quad \mu = e q,\\
&{\mathbf J}^2|\psi \rangle = j(j+1) |\psi \rangle.
\end{align}
Making use of these relations in  Eq. (\ref{eq01}) we obtain the radial Schr\"odinger equation. 

Spin formulation of the theory, given by (\ref{EQ1}), leads  to the canonical momentum, $\mathbf P = -i\nabla - e\mathbf A$, which involves the non-Abelian vector potential \cite{MAK}
\begin{align}
 e\mathbf A =- \frac{{\mathbf S}\times {\mathbf r}}{r^2}.
 \label{EqA}
\end{align}
The magnetic field strength now is determined by
\begin{align}
\mathbf B =\nabla\times \mathbf A  - i e {\mathbf A} \times {\mathbf A} = -\frac{({\mathbf S}\cdot\hat{\mathbf r})}{e}\frac{\hat{\mathbf r}}{r^2}.
\end{align}
The Abelian approach is recovered employing the unitary transformation \cite{MAK},
\begin{align}
U= \exp(-i\varphi S_3)\exp(i\theta S_2)\exp(i\varphi S_3).
\end{align}
Under this transformation, \eqref{EqA} takes the form
\begin{align}
	 e\mathbf A = -S_3 \frac{(1 - \cos \theta)}{r \sin \theta} {\mathbf e}_\varphi.
\end{align}
For the eigenvalue $-eq$ of the spin operator, this yields the Dirac vector potential \eqref{d_str} with $\mathbf n = (0,0,-1)$.

\section{Geometric phases and  artificial magnetic monopoles }
\label{GP}

Let $\mathcal H\cong\mathbb{C}^{n+1}$ be the  $(n+1)$-dimensional Hilbert space and a set $\{|e_\alpha\rangle\}, \,(\alpha=0,1,\dots,n)$ forms the orthonormal basis in $\mathcal H$.  Using the Einstein summation convention,  a given quantum state $|\psi \rangle \in \mathcal H$ may be written as $|\psi \rangle ={ Z}^\alpha |e_\alpha\rangle$, where the repeated index  $\alpha=0, \dots, n$. The space of rays is defined as an equivalence class of states $|\psi \rangle \simeq |\psi \rangle$, if $|\psi \rangle = c |\psi \rangle$, where $c \in \mathbb C$ is the complex constant. Thus, the space of rays being  the projective Hilbert space is isomorphic to the complex  projective space ${\mathbb C}P^n \cong  S^{2n+1}/U(1)$.  The normalized states belong to the unit sphere $S^{2n+1}\subset \mathbb{C}^{n+1}$.  Indeed, the normaliztion condition $ \langle\psi |\psi \rangle =1$ implies ${\bar Z}_\alpha { Z}^\alpha = 1$, where ${\bar Z}_\alpha = \delta_{\alpha\beta }{ Z}^\beta$, and $\delta_{\alpha\beta }= {\rm diag}(1,\dots,1)$. The complex projective space may be parametrized as follows: Let us assume that  $Z^0 \neq 0$, then one can introduce the coordinates in ${\mathbb C}P^n$ as $w^i= Z^i/Z^0$ $(i= 1,2,\dots,n)$.

For an arbitrary quantum evolution $t \longrightarrow|\psi (t)\rangle \in \mathcal{H}$, the geometric phase $\gamma$ gained by the system on the time interval $(0,T)$ can be written as $\gamma  = \gamma_t -\gamma_d$, where the total phase is $\gamma_t = \arg \langle \psi(0)|\psi(T)\rangle$, the dynamical phase being $\gamma_d = -i\int_0^T\langle \psi(t)|\dot{\psi}(t)\rangle dt$. 
If the trajectory is closed curve $\mathcal C$ in the complex projective space ${\mathbb C}P^n$, then the initial normalized state returns to itself up to a phase factor: $|\psi (T)\rangle = e^{i\delta(T)}|\psi (0)\rangle$.  Then the geometric phase is given by the  integral over the ${\mathcal C}\in {\mathbb C}P^n $ in the complex projective space \cite{PD}:
\begin{align}\label{Eq5}
\gamma = \oint_{\mathcal C}A \mod 2\pi,
\end{align}
where
\begin{align}\label{eq5a}
A = \frac{i}{2} \bigg ( \frac{{\bar w}_i d w^i - w_i d {\bar w}^i }{1+ {\bar w}_i  w^i }\bigg ),
\end{align}
is the connection one-form.   

Applying the Stokes's theorem, one can obtain
\begin{align}\label{eq6}
\gamma = \int_{\Sigma}F,
\end{align}
 where the integration is performed over the surface $\Sigma$ subtended by the circuit $\mathcal C$, and $F= dA$ is the curvature two-form 
\begin{align}\label{eq7}
F =  {i} \frac{{\bar w}_i w_j- ( 1 +{\bar w}_k w^k)\delta_{ij}}{(1+ {\bar w}_k  w^k )^2} d w^i  \wedge d{\bar w}^j
\end{align}
From the geometric point of view,  the geometric phase  $\gamma$ generates the holonomy element $\Gamma(\mathcal C) ={\mathcal P} \exp(i\oint_{\mathcal C}A)$, where ${\mathcal P} $ denotes path ordering,  with respect to the natural connection in a principal $U(1)$-bundle \cite{SB,Chj}. The simplest illustrative example of geometric phase is a two-level system. In this case, the ray space is the Poincare sphere $S^2$; and the geometric phase space brings out to be one-half of the solid angle 
subtended by the closed curve in ray space.

The Aharonov-Anandan (AA) phase \cite{AA},  is a particular case of the geometric phase $(\ref{Eq5})$, when the evolution of the quantum system is defined by the Schr\"odinger eqution,
\begin{eqnarray}\label{S1}
i\frac{\partial }{\partial t}|\Psi(t)\rangle = H(t)|\Psi(t)\rangle.
\label{S2}
\end{eqnarray}
Suppose that the wavefunction $|\Psi(t)\rangle$, being a solution of Eq. (\ref{S1}), satisfies the following condition
\begin{eqnarray}\label{Eq25}
|\Psi(T)\rangle = \exp(i\varphi)|\Psi(0)\rangle.
\end{eqnarray}
Let us consider a modified wavefunction 
\begin{eqnarray}\label{Eq26}
|\chi(t)\rangle = \exp(if(t))|\Psi(t)\rangle,
\end{eqnarray}
where $f(t)$ is any function satisfying $f(t+T) - f(t)= \varphi(t)$. Then, the total
phase $\varphi$ calculated for the time interval $(0,T)$ may be written as
$\varphi = \gamma + \delta$, where the  ``dynamical phase'' is given by
\begin{equation}\label{Eq27h}
 \delta = -\int_0^T \langle\chi(t)|H(t)|\chi(t)\rangle dt,
\end{equation}
and for the geometric AA phase $\gamma$ one has \cite{AA}
\begin{equation}\label{Eq27}
 \gamma = i\int_0^T \langle\chi(t)|\frac{\partial}{\partial
 t}\chi(t)\rangle dt.
\end{equation}
This yields the connection one-form and the curvature two-form as follows
\cite{AS}:
\begin{equation}\label{Eq11}
A= i\langle\chi|d\chi\rangle, \quad F=dA.
\end{equation}

Geometric phase $\gamma$ for an arbitrary quantum evolution can be obtained, also, from the total phase $\gamma_t$ by subtracting the dynamical phase $\gamma_d$ \cite{MMSY,MS,MS1,MS2}:
\begin{equation}\label{GP3}
\gamma= \gamma_t -\gamma_d,
\end{equation}
where $\gamma_t = \arg\langle \Psi(0)|\Psi(t) \rangle$, and
\begin{eqnarray}
  \gamma_d = -i \int_0^t \langle \Psi(t)|\frac{d}{dt}|\Psi(t) \rangle dt .
\end{eqnarray}

For the adiabatic evolution with the periodic Hamiltonian, $H(t+T) = H(t)$, the AA geometric phase reduces to the Berry phase as follows. Let $|\psi_n(X)\rangle$ be eigenstate corresponding to the eigenvalue $E_n$. Then in the adiabatic approximation the geometric phase is given by \cite{B1,GW,B2}
\begin{eqnarray}\label{Eq10}
 \gamma_n = i\oint_C {\langle\psi_n(X)|d\psi_n(X)\rangle}.
\end{eqnarray}
Here  we assume that the instantaneous eigenvectors form the orthonormal basis, $\langle\psi_m|\psi_n\rangle =\delta_{mn}$. 
 
Validity of the adiabatic approximation is defined by the following condition:
\begin{equation}\label{Eq24}
    \sum_{m\neq n}\bigg|\frac{\langle\psi_m|\partial H/\partial
    t|\psi_n(X)\rangle}{(E_m - E_n)^2}\bigg|\ll 1
\end{equation}
This restriction is violated nearby the degeneracies, where the eigenvalues coalesce.

Inserting in  the expression $F^{n} = dA_{n} $ the complete basis system condition, $ \sum_m|\psi_m\rangle \langle \psi_m|= {1\hspace{-0.15cm}1} $,  and employing the Schr\"odinger equation, $H|\psi_m \rangle = E_m |\psi_m\rangle$, we obtain
\begin{align}
\label{B0a}
F^{n}= -i\sum^N_{m\neq n}\frac {\langle  \psi_n|\nabla_a H |\psi_m\rangle
 \langle \psi_m| \nabla_b H |\psi_n\rangle d\lambda^a\wedge d\lambda^b } {(E_m-E_n)^2}
  \end{align}
It follows herefrom that the curvature $F^{(n)}$ diverges at the degeneracy points $\lambda^*$, where the energy levels, say $E_n$ and $E_{n+1}$, are crossing, $E_n(\lambda^*) =E_{n+1}(\lambda^*)$.
For further it is of importance the following result: $ \sum_n F^{n} =0$ \cite{Chj}.

\subsection{Diabolical points and ``magnetic'' monopoles}

\begin{figure}[tbh]
\scalebox{0.45}{\includegraphics{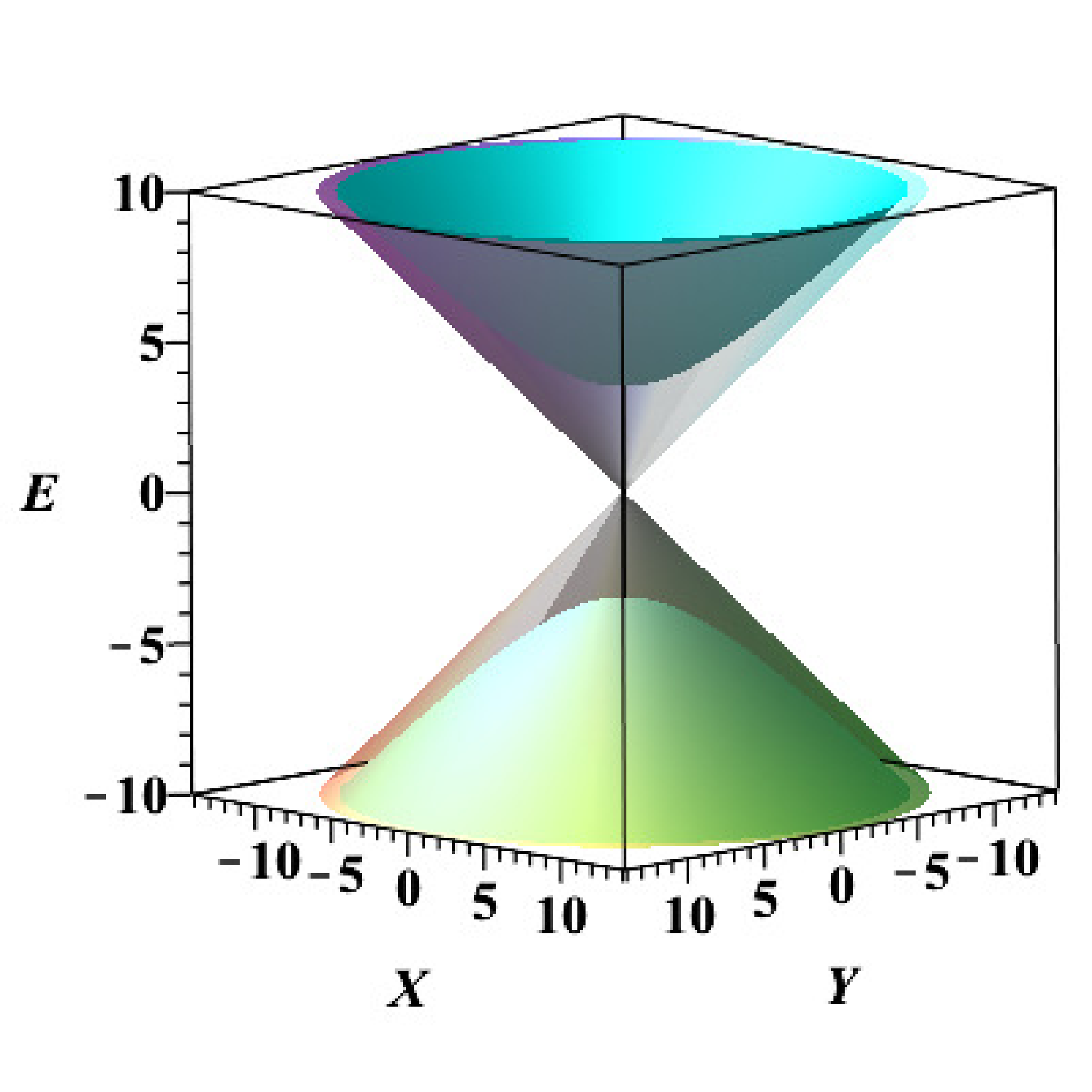}}
\caption{Dependence of the energy $E_{\pm}$ on ($X,Y$) for fixed values of $Z$ ($\lambda_0 =0$). Energy surfaces avoiding crossing ($Z=5$). Energy surfaces crossing occurs at the DP ($X=Y=Z=0$).
\label{DP}}
\end{figure}

In the typical case of double degeneracy with two linearly independent eigenvectors, the energy surfaces form the sheets of a double cone (see Fig. \ref{DP}). The apex of the cones is called a ``diabolical point'' (DP) \cite{B0,BW}.  For a generic Hermitian Hamiltonian, the co-dimension of the DP is three,  and therefore, it can be characterized by three parameters: $\mathbf R= (X,Y,Z)$ \cite{Arn}. The eigenstates, $|n,\mathbf R \rangle$, give rise to the Berry's connection defined as ${\mathbf A}_n(\mathbf R)= i\langle n,\mathbf R| \nabla_{\mathbf R} |n,\mathbf R \rangle$, and the curvature,
$\mathbf B_n = \nabla_{\mathbf R} \times {\mathbf A}_n $, associated
with ${\mathbf A}_n$ is the field strength  of `magnetic' monopole located at
the DP \cite{B0,B1,BW,BD}. The Berry phase, $\gamma_n= \oint_{\mathcal C }{\mathbf A}_n \cdot d \mathbf R$, is interpreted as a holonomy associated with parallel transport along a circuit $\mathcal C$ \cite{SB}.

Since in the vicinity of each DP only terms related to the invariant subspace formed by the two-dimensional Jordan block make substantial contributions, the $N$-dimensional problem becomes effectively two-dimensional (for details see \cite{Arn,KMS}). This implies that there exists the map $\varphi :\mathfrak M \mapsto S^2$ such that in the vicinity of the DP the quantum system can be described by the effective two-dimensional Hamiltonian, $H_{ef}= (1/2)\lambda_0 {1\hspace{-.15cm}1}+(1/2)\mathbf R\cdot \boldsymbol \sigma$, where $R= |E_{n+1} - E_n|$. This yields $\gamma_n \approx \int_{\Sigma^\prime} F^{n}$, where $\Sigma^\prime=\varphi (\Sigma) \subset S^2$ and
\begin{align}\label{B7}
F^{n} =q_n\frac{\mathbf R \cdot d\mathbf S }{R^3} .
\end{align}
Thus, the geometric phase is independent of peculiarities of a quantum-mechanical system.  

In what follows, we consider in detail the  Berry phase associated with a generic Hamiltonian of two-level system:
\begin{equation}\label{eqH2a}
H=\frac{1}{2}\left(
  \begin{array}{cc}
    \lambda_0 + Z & X-iY \\
   X+iY & \lambda_0 - Z\\
  \end{array}
\right), \quad X,Y,Z \in \mathbb R.
 \end{equation}
The solution of the eigenvalue problem, $H|u\rangle = E|u\rangle$, is given by
$E_{\pm} = \lambda_0/2 \pm R/2$,  where $R = {(X^2 +Y^2 + Z^2)}^{1/2}$.  We choose the eigenvectors as,
\begin{eqnarray}
|u_{+}\rangle =& \left(\begin{array}{c}
                  \cos\frac{\theta}{2}\\
                  e^{i\varphi}\sin\frac{\theta}{2}
                  \end{array}\right),
 \label{r}\\
|u_{-}\rangle =& \left(\begin{array}{c}
-e^{-i\varphi}\sin\frac{\theta}{2}\\
\cos \frac{\theta}{2} \end{array}\right), \label{l}
\end{eqnarray}
where $(R,\theta, \varphi)$ are the spherical coordinates in the parameter space. In the basis of the eigenstates $|u_{\pm}\rangle$ the Hamiltonianan (\ref{eqH2a}) reads
\begin{align}\label{H1a}
H= \frac{\lambda_0 }{2}{1\hspace{-0.15cm}1} + {R} S_3,
\end{align}
where $S_3 = \sigma_3/2$.

As one can see, the coupling of eigenvalues $E_{+}$ and
$E_{-}$ occurs at the DP, determined by the equation
\begin{equation}\label{Eq9}
X^2 +Y^2 + Z^2=0.
\end{equation}
Thus, the DP  is located at the origin of coordinates in 3-dimensional parameter space, $\mathbb{R}^3$. At the DP we obtain
\begin{eqnarray}
|u_{+}\rangle = \left(\begin{array}{r}
                  1\\
                  0
                  \end{array}\right),\quad
|u_{-}\rangle = \left(\begin{array}{c}
0\\
1 \end{array}\right). \label{left}
\end{eqnarray}

Inserting formulae (\ref{r}) and (\ref{l})  into Eq. (\ref{Eq11}), we obtain the connection one-form as,
\begin{eqnarray}\label{Eq12}
    A_{\pm}=  q_{\pm}(1 -\cos\theta)d\varphi ,
\end{eqnarray}
where $q_{\pm}= \mp 1/2$, and  upper/lower sign corresponds to $|u_{\pm}\rangle$, respectively. The related curvature two-form reads
\begin{eqnarray}\label{Eq14}
 F_{\pm}=dA_{\pm}=q_{\pm}\sin\theta\; d\theta \wedge d\varphi, \; \theta, \varphi \in \mathbb C,
\end{eqnarray}
and computation of the geometric phase yields
\begin{equation}\label{GP1}
\gamma_{\pm} = \oint_{ C} A_{\pm},
\end{equation}
where integration is performed over the contour $\mathcal C$ on the
sphere $S^2$. Applying the Stokes theorem we obtain
\begin{equation}\label{GP2}
\gamma_{\pm} = \int_{\Sigma} F_{\pm} = q_{\pm}\Omega(\mathcal C),
\end{equation}
where $\Sigma$ is a closed surface with the boundary $\mathcal C= \partial \Sigma $,
and $\Omega(\mathcal C)$ is the solid angle subtended by the contour $\mathcal C$.

In the Cartesian coordinates the  connection one-form and the curvature two-form can be written as,
\begin{align}\label{Eq16}
A_{\pm} =& \frac{ q_{\pm}(XdY-YdX)}{R(R + Z)} ,\\
F_{\pm} = &\frac{q_{\pm}}{R^3}\varepsilon_{ijk} X^k dX^i\wedge dX^j.
\label{Eq16a}
\end{align}

The obtained formulae describe two artificial `magnetic' monopoles with the charges $q_{+}$ and  $q_{-}$. One can write the field strength of these monopoles as
\begin{equation}\label{Eq4a}
\mathbf B = q\frac{\mathbf R}{R^3},
\end{equation}
where $q=q_{\pm}$ and $\mathbf R = (X,Y,Z)$. 

In general case, the vector potential involved in Eq. (\ref{Eq16}) can be recast as,
\begin{equation}
{\mathbf A}= q\frac{{\mathbf R}\times {\mathbf n}}
{R(R - {\mathbf n} \cdot{\mathbf R})},
\label{Eq17d}
\end{equation}
where the unit vector $\mathbf n$ determines the direction of a  singular string passing from the origin of coordinates to infinity.  For particular choice of $\mathbf n = (0,0,-1)$ we obtain the vector potential (\ref{Eq16}).

\subsection{Spin of artificial monopole from isospin} 

The phenomen of conversion of isopspin degrees of freedom into the spin degrees of freedom was discovered in the $\rm SU(2)$  quantum gauge field theory, with the spontaneuosly broken isospin symmetry, in the field of magnetic monopole \cite{HPTHG,MSW,RJRC}.  It was demonstrated that the similar phenomenon can arise in some quantum-mechanical systems with induced non-Abelian gauge fields, when there exists isotopic-spin-degenerate energy levels \cite{LHZ,LHZ1}. 

For the monopole (\ref{Eq4a})  we introduce an isospin assuming that the internal (isotopic) degrees of freedom are related with the energy levels crossing at the DP.
Below we show that isospin contributes to the total angular momentum and acts as a spin, confirming that the monopole spin comes out from isospin.

Consider a non-Abelian gauge field with  internal (isotopic) degrees of freedom:
${\mathbf A} \rightarrow {\mathbf A} = {\mathbf A}^a I_a $,
where $I_a = \tau_a/2$  are the generators of isospin, $\tau_a$ being the Pauli matrices.  
Assuming ${\mathbf A}^a ={\mathbf A}_D\delta^3_a $, where ${\mathbf A}_D$ is  taking from Eq. (\ref{Eq12}),
\begin{align}
A^r_D=0, \quad A^\theta_D =0, \quad  A^\varphi_D =\frac{1-\cos \theta}{R \sin \theta},
\end{align}
we obtain $ {\mathbf A} = -{\mathbf A}_D  I_3$. The associated field strength is the ``magnetic'' field of the monopole,
\begin{equation}\label{Eq4k}
{\mathbf B }=- I_3\frac{\mathbf {\hat R}}{R^2} ,
\end{equation}
 where $\hat {\mathbf R}= {\mathbf R}/R$. Thus, the different charges of the monopole (\ref{Eq4a}), being associated with  $E_{\pm}$ energy surfaces crossing at the DP, can be described as the monopole with isotopic degrees of freedom.

The total angular momentum, $ {\mathbf J } =\mathbf R \times (-i\nabla- {\mathbf A} ) +\hat {\mathbf R}I_3$, yields the following commutation relations:
\begin{eqnarray}
[J_i, J_j] = i\epsilon_{ijk}J_k.
\end{eqnarray}
One can show that ${\mathbf J }$ commutes with the Hamiltonian (\ref{H1a}), and, thus, ${\mathbf J }$ is conserved.

While explicitly the vector potential ${\mathbf A}$ does not manifest itself as a rotational invariant field, it is rotationally symmetric in the sense that there exists a gauge transformation ${\mathbf A} \rightarrow U^{-1} {\mathbf A} U + i U^{-1}\nabla U$ that compensates the rotational noninvariance of ${\mathbf A}$  \cite{JR1}. 

Consider the following gauge transformations rotating
the unit vector $\hat{\mathbf R}$ on the sphere $S^2$:
\begin{align}
U(\theta,\varphi)=\left(
\begin{array}{cc}
\cos\frac{\theta}{2} e ^{i\varphi/2}& \sin\frac{\theta}{2}e ^{-i\varphi/2}\\ 
-\sin\frac{\theta}{2}e ^{i\varphi/2}& \cos\frac{\theta}{2}e ^{-i\varphi/2}
\end{array} 
\right ).
\end{align}
As one can see,  $U(\theta,\varphi)$ brings the Hamiltonian (\ref{H1a}) to its initial form (\ref{eqH2a}):
\begin{align}\label{H1b}
H= \frac{\lambda_0 }{2}  {1\hspace{-0.15cm}1} +{\mathbf R}\cdot{\mathbf S}.
\end{align}
Here ${\mathbf S} ={\bm \sigma}/2$ comprises the Pauli matrices. 
In a new gauge,  after some work one obtains \cite{JR1,Hooft,PAM,AJFP}
\begin{align}\label{JA1}
{\mathbf A} =& \frac{{\mathbf R}\times {\mathbf S}}{R^2}, \\
{\mathbf B }= &-\frac{ {\mathbf S} \cdot \hat{\mathbf R} }{R^3}  {\mathbf R}, \label{B1} \\
{\mathbf J } =& \mathbf R \times (-i\nabla - {\mathbf A} ) + \hat{\mathbf R }\, ({\mathbf S} \cdot \hat{\mathbf R}).
\label{J1a}
\end{align}

The curvature associated with the gauge field $ {\mathbf A}$ is:
\begin{align}
{ F}_{ij} =\partial_i {A}_j - \partial_j { A}_i -i[{ A}_i,{A}_j] , \quad
{ B}_i = (1/2)\varepsilon_{ijk} { F}_{jk} .
\end{align}

 Next, inserting (\ref{JA1}) into Eq. (\ref{J1}), we obtain ${\mathbf J } = {\mathbf L }  + {\mathbf S } $,
 where ${\mathbf L } = \mathbf R \times \mathbf P$ is a standard operator of the angular momentum.
 Thus, we find that  the total angular momentum  $\mathbf J $  and projection of spin on direction $\mathbf R$ are conserved.  Quantization is performed in the unitary gauge, where $\mathbf R$ is pointed in the third isospin direction:
\begin{align} \label{S1a}
& ({\mathbf S} \cdot \hat{\mathbf R} )\,\psi = m\psi, \\
&H\psi = E\psi.
\end{align}

Evolution of the monopole spin is determined by the following equations of motion: 
\begin{align}
\dot{\mathbf S} = i[H,{\mathbf S}],
\end{align}
For the Hamiltonian (\ref{H1b}) we obtain
\begin{align}
\dot{\mathbf S}=  {\mathbf S}\times {\mathbf R},
\end{align}
which is equivalent to the Bloch equation.
  
\section{Quasienergies and geometric phases for time-periodic quantum systems} 
\label{QE}

\subsection{Quasienergies and monopole spin}

When a quantum $N$-level system interacts with the periodic external field, a dressed state (or quasi-particle) appears. This quasi-particle is characterized by a combination of the parameters of the system and the external field. In particular, not energy but quasienergy becomes a good quantum number \cite{RVI,BPXH}. Below we will show that this quasi-particle gains additional degrees of freedom related to the artificial magnetic monopole at the DP's vicinity. Thus, this monopole contributes to the observables, such as the quasienergy, susceptibility, and others.

Consider a time-periodic Hamiltonian $H(t)$ with the period $T = 2\pi/\omega$, such that $ H(t + T) =  H(t)$. Let $\vartheta = \omega t$ be a new variable. Then one can introduce the  time-independent Floquet Hamiltonian, 
\begin{align}
{H}_F = H - i\omega\partial/\partial \vartheta,
\end{align}
 acting in the extended Hilbert space $\mathcal H \otimes \mathcal F$, where $\mathcal F = L^2(S^1,d\vartheta/2\pi)$ is the space of square integrable functions on the circle $S^1$. The extended Hilbert space  $\mathcal H \otimes \mathcal F$ is equipped by the scalar product: 
$$\langle\langle \phi|\psi \rangle \rangle = \int_0^{2\pi}\langle \phi|\psi  \rangle d\vartheta/2\pi.$$

Employing the Floquet theorem \cite{SHI,HGM}, one can show that the solution of the Schr\"odinger equation can be written as,
 \begin{align}\label{Eq22}
|\Psi_{\varepsilon}(t) \rangle = e^{-i\varepsilon t} |\Phi_{\varepsilon} (t) \rangle,
\end{align}
where $|\Phi_{\varepsilon}(t) \rangle$ is the periodic wafefunction with the period $T$.
The quasienergies $\varepsilon$ are defined from the eigenvalue problem for the Floquet Hamiltonian:
\begin{align}
{ H}_F|\Phi_{\varepsilon}(\vartheta) \rangle = \varepsilon |\Phi_{\varepsilon} (\vartheta)\rangle.
\label{EF}
\end{align}

The total phase accumulated by the quantum  system in the period $T$ is given by
$\gamma_t = \arg\langle \Psi_\varepsilon(0)|\Psi_\varepsilon(T) \rangle= - \varepsilon T$, and the computation of the dynamical phase, 
\begin{eqnarray}
\gamma_d = -i\int_0^T \langle\Psi_\varepsilon (t)|\dot\Psi_\varepsilon (t)\rangle dt ,  
\end{eqnarray}
yields the following result:
\begin{eqnarray}\label{Eq23}
\gamma_d =-\varepsilon T  -i\int_0^T \langle\Phi_\varepsilon(t)|\dot\Phi_\varepsilon (t)\rangle dt.
 \end{eqnarray}
From here we obtain the geometric phase  $\gamma_\varepsilon= \gamma_t - \gamma_d$ of the state $|\Phi_\varepsilon \rangle $ as \cite{MDJ,MDJ1,MDJ2},
\begin{eqnarray}\label{Eq23b}
\gamma_{\varepsilon} = i\int^{2\pi}_0  \langle \Phi_{\varepsilon}(\vartheta) |\partial\Phi_{\varepsilon} (\vartheta)/\partial \vartheta\rangle d \vartheta.
 \end{eqnarray}
Next, one can apply the Hellmann-Feynman theorem,
\begin{eqnarray} \label{Eq31}
 \frac{\partial \varepsilon}{\partial \omega}& = &\frac{1}{2\pi}\int_0^{2\pi} \bigg \langle\Phi_\varepsilon(\vartheta)\bigg|\frac{\partial  H_F}{\partial \omega}\bigg|\Phi_\varepsilon(\vartheta)\bigg\rangle d \vartheta\nonumber \\
 & = &- \frac{i}{2\pi}\int_0^{2\pi} \langle\Phi_\varepsilon(\vartheta)|\frac{\partial}{\partial\vartheta }\Phi_\varepsilon(\vartheta)\rangle d\vartheta,
\end{eqnarray}
to obtain a simple formula for the geometric phase accumulated by the quantum system in the period \cite{GMPH},
\begin{eqnarray}\label{Eq28}
\gamma_\varepsilon = -2\pi \frac{\partial \varepsilon}{\partial \omega}.
\end{eqnarray}

As an illustrative example, let us consider a quantum mechanical system with double degeneracy of quasienergy occurring at the DP. Near the DP the system is characterized by three parameters, $\mathbf R = (X,Y,Z)$, and its dynamics can be described by effective Hamiltonian (\ref{H1b}).

 Returning back to the geometric phase  of a Floquet state, we find that Eq. (\ref{Eq23b}) can be recast as, $\gamma_{\varepsilon}=  \oint_{\mathcal C} \mathbf A \cdot d\mathbf R$, where
 \begin{align}
 	 \mathbf A \cdot d\mathbf R = i\langle\Phi_\varepsilon \big (\mathbf R(\vartheta) \big )|\frac{\partial}{\partial\vartheta }\Phi_\varepsilon\big (\mathbf R(\vartheta) \big )\rangle d\vartheta
 \end{align}
 is the connection one-form, and the integral is taken over the circuit $\mathcal C  \subset S^2$. To recover information on the monopole structure hidden in the quasienergy, we employ Eq. \eqref{B1} to obtain 
\begin{equation}\label{GP2a}
2 \pi \frac{\partial \varepsilon}{\partial \omega} = \Omega(\mathcal C)\langle{\mathbf S} \cdot \hat{\mathbf R} \rangle  \mod (2 \pi),
\end{equation}
where $\mathbf S$ is the monopole spin, $\langle{\mathbf S} \cdot \hat{\mathbf R} \rangle=\langle \Phi_\varepsilon |{\mathbf S} \cdot \hat{\mathbf R} |\Phi_\varepsilon \rangle$, and $\Omega(\mathcal C)$ is the solid angle subtended by the contour $\mathcal C$.

 Suppose that the Hamiltonian of the system depends on the perturbation parameter $\lambda$ in such a away that for adiabatic switch of interaction  $H(\lambda,t)\stackrel{\lambda\rightarrow 0}{\longrightarrow}\hat H_0$, where $\hat H_0$ is time-independent Hamiltonian. Then the quasienergies satisfy \cite{GMPH}
\begin{align}\label{eq1}
\lim_{\lambda\rightarrow 0}\varepsilon_n(\omega,\lambda)= \varepsilon_{n}(\omega) = E^0_n + p\omega, \, p\in \mathbb{Z},
\end{align}
where we denote by $E^0_n$ the eigenenergies of the unperturbated Hamiltonian $H_0$.
From here and Eq. (\ref{Eq28}) we find that in the adiabatic limit of vanishing perturbation $(\lambda\rightarrow 0)$ the geometric phase takes the trivial value of $\gamma_{\varepsilon}= -2\pi p$.

Let us assume that $H(t) = H_0 +\lambda V(\omega t)$, where $H_0$ is `unperturbated' stationary Hamiltonian and $V(\omega t)$ is a periodic operator. By differentiating the eigenvalue equation $\hat{ H}_F|\Phi_{\varepsilon_n} \rangle = \varepsilon_n |\Phi_{\varepsilon_n} \rangle$, one obtains the relations \cite{AK,BWB,SJH}
\begin{align}\label{eq9}
\langle \langle\Phi_{\varepsilon_n}|\lambda V|\Phi_{\varepsilon_n} \rangle \rangle = \lambda\frac{\partial \varepsilon_n}{\partial \lambda}.
\end{align}
Considering the perturbation parameter $\lambda$ as  the `field' strength, one may introduce the generalized nonlinear susceptibility as,
\begin{align}\label{eq10}
\chi_n(\omega,\lambda)= \frac{\partial \varepsilon_n}{\partial \lambda}.
\end{align}
In the limit $\lambda \rightarrow 0$ it leads to (up to the numerical constant) widely known in the literature linear susceptibility: 
$\chi_n(\omega,\lambda) \rightarrow \chi_n(\omega).$

Let us assume that 
\begin{align}
H  = \frac{1}{2}
\left(\begin{matrix}
  E_1&0\\
  0&E_2 \\
\end{matrix}
\right) +
\frac{\lambda}{2}
\left(\begin{matrix}
  0&V_0 e^{-i\omega t} \\
 V_0^\ast e^{i\omega t} &0 \\
\end{matrix}
\right)
\end{align}
It yields the following Floquet Hamiltonian, 
\begin{align}
{H}_F = E_0 \one + \frac{1}{2}
\left(\begin{matrix}
  \omega_0&\lambda V_0 e^{-i\vartheta} \\
 \lambda V_0^\ast e^{i\vartheta} &-\omega_0\\
\end{matrix}
\right)
 - i\omega\partial/\partial \vartheta,
\end{align}
where $E_0 = (E_1 + E_2)/2$ and $\omega_0 = E_1 - E_2 $. The quasienergies are defined from the eigenvalue problem for the Floquet Hamiltonian:
\begin{align}
{ H}_F|\Phi_{\varepsilon}(\vartheta) \rangle = \varepsilon |\Phi_{\varepsilon} (\vartheta)\rangle.
\label{EF1}
\end{align}
We obtain
\begin{align}
	\varepsilon^p_{\pm} = E_0 +p \omega +\frac{1}{2} \Big ( \omega \pm \sqrt{\lambda^2|V_0|^2 + \delta ^2}\Big ), \quad p \in \mathbb Z ,
	\label{EV3}
\end{align}
where $\delta = \omega_0 - \omega$ is the detuning. The corresponding eigenvectors are
\begin{align}
	|u^p_{+}\rangle =& e^{ip\vartheta}\left(\begin{array}{c}
                  \cos\frac{\theta}{2}\\
                  e^{i\vartheta}\sin\frac{\theta}{2}
                  \end{array}\right),
 \label{r_1}\\
|u^p_{-}\rangle =& e^{ip\vartheta}\left(\begin{array}{c}
-e^{-i\vartheta}\sin\frac{\theta}{2}\\
\cos \frac{\theta}{2} \end{array}\right), 
\label{l_1}
\end{align}
where $(\vartheta, \theta)$ are the spherical angles in the parameter space:
\begin{align}
X &=R \sin \theta \cos \vartheta, \\ 	
Y &=R \sin \theta \sin \vartheta, \\
Z&=R \cos\theta,
\end{align}
where $R =\sqrt{\lambda^2|V_0|^2  + (\omega_0 - \omega)^2}$ and $Z = \omega_0 - \omega$. 

First, using \eqref{Eq23b}, we calculate the geometric phase for the contour $\theta = \text {const}$, 
\begin{align}
\gamma_{\pm} =  \mp\pi(1 - \cos \theta) =  \mp\frac{1}{2}\Omega(\mathcal C),
\label{GP1a}
\end{align}
where $\Omega(\mathcal C)$ is the solid angle subtended by the contour  $\theta = \text {const}$. Taking the derivative of the quasienergy with respect to the frequency, $\omega$, we obtain
\begin{align}
 2\pi \frac{\partial \varepsilon^p_{+}}{\partial \omega} &=  \pi(1 - \cos \theta) +2\pi p , \\
 2\pi \frac{\partial \varepsilon^p_{-}}{\partial \omega}& = -\pi(1 - \cos \theta) +2\pi (p +1).
\end{align}
Comparison of these results with \eqref{GP1} yields
\begin{align}
2\pi \frac{\partial \varepsilon_{\pm}}{\partial \omega} = - \gamma_{\pm }  \mod(2\pi ),
\label{GP3b}
\end{align}
that is in agreement with Eq. \eqref{Eq28}. 
 \begin{figure}[tbh]
\scalebox{0.35}{\includegraphics{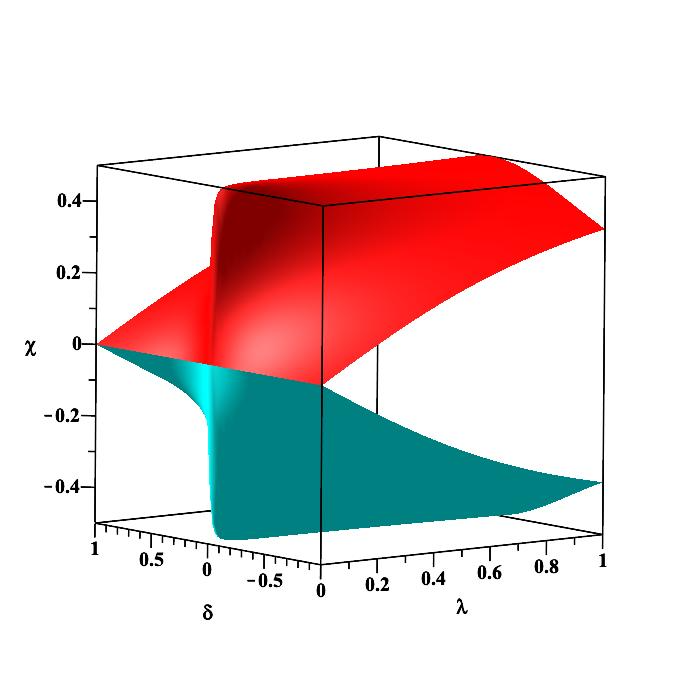}}
\caption{Susceptibility, $\chi$,  as a function of the detuning, $\delta= \omega_0 - \omega$, 
and $\lambda$; $\chi_+ $ (red (upper) surface), $\chi_- $ (cyan (lower) surface).
\label{chi1}}
\end{figure}

 Next, using  Eqs.\eqref{Eq28}, \eqref{GP2a} and  \eqref{GP1}, we obtain
\begin{equation}\label{GP3a}
\Omega(\mathcal C)\langle{\mathbf S} \cdot \hat{\mathbf R} \rangle = \pm\frac{1}{2}\Omega(\mathcal C) .
\end{equation}
From here it follows $\langle{\mathbf S} \cdot \hat{\mathbf R} \rangle = \pm 1/2$. Thus, we conclude that with the quasienergy $\varepsilon_{\pm}$ is associated the projection of the monopole spin $m =\pm 1/2$.

The computation of the generalized nonlinear susceptibility yields
\begin{align}\label{eq10d}
\chi_{\pm}= \pm\frac{\lambda |V_0|^2 }{2\sqrt{\lambda ^2 |V_0|^2 + (\omega_0 - \omega)^2}} .
\end{align}
In Fig. \ref{chi1} we depict the nonlinear susceptibility, $\chi$, as a function of $\lambda$ and the detuning $\delta$. As one can see, $\lambda =0 $ is the bifurcation line.

 Let us assume that the perturbation switch on adiabatically, then the susceptibility takes the value $\chi_+$ or $\chi_-$ as a result of the spontaneous symmetry breaking. As a consequence, the monopole spin's projection on the quantization axes takes the value $1/2$ or  $-1/2$. Thus, an experiment with measuring nonlinear susceptibility offers the possibility of the experimental test for the monopole spin.

\subsection { Superconducting qubit interacting with a resonator} 

As the following illustrative example, we consider a superconducting qubit system interacting with a resonator. In the rotating wave approximation, the system is described by the Hamiltonian
\begin{eqnarray}\label{Eq32}
H= \frac{\omega_q}{2} \sigma_z + \omega_r a^\dagger a + \frac{\lambda}{2}(\sigma_{+} a + \sigma_{-} a^{\dagger}).
\end{eqnarray}

We study the dynamics of the system under the condition of a given field approximation for the resonator. We suppose that initially the system is prepared in the pure state
\begin{eqnarray}\label{Eq33}
|\psi(0)\rangle = |u(0)\rangle\otimes|\psi_p(0)\rangle,
\end{eqnarray}
where $|u(0)\rangle = a_{+ }|+\rangle + a_{- }|-\rangle$ is a general qubit state, with $|\pm\rangle$ being its eigenstates, and $|\psi_p(0)\rangle$ is the coherent state
\begin{eqnarray}\label{Eq34}
|\psi_p(0)\rangle = e^{-|\alpha|^2/2} \sum_{n=0}^\infty \frac{\alpha^n}{\sqrt{n!}}\, |n\rangle.
\end{eqnarray}
Further it is assumed $\alpha$ be real, that does not restrict the generality of consideration.

Assuming $\alpha \gg 1$, one can write the time-dependent state as \cite{MDJ1}
\begin{eqnarray}\label{Eq38a}
&|\psi(t)\rangle  = e^{-\alpha^2/2} \sum_{n=1}^\infty \frac{\alpha^n}{\sqrt{n!}}\bigg (C_{-,n} e^{-i\varepsilon_{-,n} t}|\Phi_{-,n}\rangle \nonumber \\
&+ C_{+,n} e^{-i\varepsilon_{+,n} t}|\Phi_{+,n}\rangle \bigg),
\end{eqnarray}
where
\begin{eqnarray}\label{Eq38b}
C_{+,n} = \frac{\alpha}{\sqrt{n+1}}\,a_{-}\beta_{+,n}- a_{+}\beta_{-,n}, \\
C_{-,n} = \frac{\alpha}{\sqrt{n+1}}\,a_{-}\beta_{-,n} + a_{+}\beta_{+,n}.
\end{eqnarray}
Here
\begin{align}
\beta_{\pm,n}= \sqrt{(\Omega_n \pm \omega_q \mp \omega_r)/(2\Omega_n)}
\end{align}
and 
\begin{align}
\Omega_n = \sqrt{(n+1)\lambda^2 + (\omega_q -\omega_r)^2}. 
\end{align}

As can be seen the states $|\Psi_{\pm,n}\rangle$ defined as
\begin{eqnarray}
|\Psi_{\pm,n}\rangle = e^{- i\varepsilon_{\pm,n} t}|\Phi_{\pm,n}\rangle
\end{eqnarray}
are the Floquet states. Indeed, as can be easily shown, the Floquet modes $|\Phi_{\pm,n}\rangle$ satisfies
\begin{equation}\label{Eq38c}
 H_F|\Phi_{\pm,n}\rangle = \varepsilon_{\pm,n}|\Phi_{\pm,n}\rangle,
\end{equation}
where
\begin{equation}\label{Eq39}
 \varepsilon_{\pm,n} =\bigg (n + \frac{1}{2}\bigg)\omega_r \mp \frac{\Omega_n}{2}
\end{equation}
is the quasienergy.

To  calculate the geometric phase we apply  Eq. (\ref{Eq28}) written as,
\begin{eqnarray}
\gamma_{\pm,n} = -2\pi \frac{\partial \varepsilon_{\pm,n}}{\partial \omega_r}.
\end{eqnarray}
The computation yields
\begin{eqnarray}\label{Eq40}
\gamma_{\pm,n} = \pm\pi\cos\theta_n - 2\pi \Big(n+\frac{1}{2}\Big),
\end{eqnarray}
where $\cos \theta_n = (\omega_r - \omega_q)/\Omega_n$.

As can be shown, the Floquet states $|\Phi_{\varepsilon_m} \rangle$ are eigenfunctions of the both operators $ H_F$ and ${\mathbf S}\cdot \hat {\mathbf R} $,
\begin{align}
 H_F |\Phi_{\pm,n}\rangle= \varepsilon_{\pm,n} |\Phi_{\pm,n}\rangle, \\
({\mathbf S}\cdot \hat {\mathbf R}) |\Phi_{\pm,n}\rangle = \pm \frac{1}{2}|\Phi_{\pm,n}\rangle.
\end{align}
Thus,  with the quasienergy $\varepsilon_{\pm,n}$ is associated the projection $\pm 1/2$ of the monopole spin.

In semiclassical limit the corresponding Hamiltonian can be obtained from the initial one by replacing $n$ in the expression for $\theta_n$ with its mean value, $\bar n =\alpha^2$, and the boson creation operator $a^\dagger$ by the function $\alpha e^{i\omega_r t}$. It takes the form \cite{MDJ1},
\begin{eqnarray}
H_{q}  = \frac{1}{2}\left(
                      \begin{array}{cc}
                        \omega_q & \kappa e^{-i\omega_r t}\\
                        \kappa e^{i\omega_r t} & - \omega_q \\
                      \end{array}
                    \right),
\end{eqnarray}
where $\kappa = \alpha\lambda$ is the effective coupling constant.

The geometric phase of (\ref{Eq40}) becomes the nonadiabatic geometric phase
\begin{equation}\label{Eq41}
 \gamma_{\pm,n} \rightarrow \gamma = -\pi(1 \mp\cos\theta)
\end{equation}
with $\cos \theta = (\omega_r - \omega_q)/\Omega$ and $\Omega= \sqrt{\kappa^2 + (\omega_q - \omega_r)^2 }$. In the adiabatic limit, $\omega_r/\sqrt{\kappa^2 + \omega_q^2 } \ll 1$, it reduces to the Berry phase
\begin{equation}\label{Eq41a}
 \gamma \rightarrow \gamma_b = -\pi\bigg(1 \mp \frac{\omega_q}{\sqrt{\kappa^2 + \omega_q^2 }}\bigg).
\end{equation}

\subsection{Spin system driven by circularly polarized field}

Let us consider the spin-$j$ system driven by a circularly polarized time-periodic field. The Hamiltonian of the system we take in the form \cite{LHCH}:
\begin{align}\label{Eq1a}
 H=  (\mu_0/j)\mathbf{B}(t)\cdot {\mathbf{J}},
 \end{align}
where $ \mu_0 $ is a magnetic momentum, and ${\mathbf{J}}$ is the spin angular momentum operator. Its components obey the standard commutation relations: $[J_i,J_j] = i\varepsilon_{ijk}J_k$.

Further we restrict ourselves by consideration of the time-dependent periodic magnetic field rotating in the $(x,y)$-plane, thus that $\mathbf B(t)= (B_\perp\cos(\omega t-\varphi),B_\perp\sin(\omega t- \varphi),B_z)$. Introducing $\omega_0 =\mu_0 B_s/j$ and $V = \mu_0 {B_\perp}e^{i\varphi}/j$, we recast the Hamiltonian of Eq.(\ref{Eq1a}) as
 \begin{align}\label{Eq1b}
 H= \omega_0  J_3 +\frac{V^\ast}{2}e^{i\omega t}{ J}_{+} 
+\frac{V}{2}e^{-i\omega t}{ J}_{-}  
\end{align}

To obtain the matrix representation of the Floquet Hamiltonian, we first introduce the Floquet state basis 
$|j,m;p\rangle = |j,m\rangle\otimes |p\rangle$, where $|j,m\rangle$ are eigenvectors of the operators ${ J}_3$ and ${ {\mathbf J}}^2= { J}^2_1+{ J}^2_2+{ J}^2_3$, so that ${ J}_3|j,m\rangle = m|j,m\rangle$ and ${ {\mathbf J}}^2|j,m\rangle = j(j+1)|j,m\rangle$ ($m=-j,-j+1, \dots, j-1,j$). 

In terms of the basis $|j,m;p\rangle$ the Floquet Hamiltonian $ H_F$ reads
\begin{align}\label{Eq1c}
&\langle j,m;p| H_F|j,n;q\rangle  =(m \omega_0 + p \omega )\delta_{mn}\delta_{pq} \nonumber \\
&+\frac{1}{2}\big( V^\ast\langle j,m|{ J}_{+}|j,n\rangle \delta_{p, q-1} 
+ V \langle j,m|{ J}_{-}|j,n\rangle \delta_{p, q+1} \big ),
\end{align}
where
\begin{align}
\langle j,m|{ J}_{\pm}|j,n\rangle = \sqrt{j(j+1) - n(n \pm 1)} \delta_{m,n\pm1}.
\end{align}

As can be observed the infinite-dimensional Floquet Hamiltonian has a block-diagonal structure with dimension of each block being $(2j+1)\times (2j+1)$. Due the periodicity of quasi-energies only one block can be considered.  We choose the central block defined by the subspace spanned by the following set of the Floquet basis $\{|j,m;-m-j\rangle \}$  $(m=-j,-j+1,\dots,j)$. Then the  Floquet Hamiltonian (\ref{Eq1c}) reduces to the $(2j+1)$-dimensional Hamiltonian
\begin{align}\label{Eq1d}
 H_j = -j\omega  {1\hspace{-0.15cm}1} + \Omega{\mathbf S}\cdot \hat {\mathbf R},
\end{align}
where  $  {\mathbf S}\cdot  {\mathbf R}= \sin\theta \cos\varphi { J}_{1} + \sin\theta \sin\varphi { J}_{2} + \cos\theta { J}_{3}$,  and $ {1\hspace{-0.15cm}1}$ is the identity operator.  We set $\cos\theta = \Delta/\Omega$, where
$\Delta = \omega_0 -\omega$ and $\Omega = \sqrt{\Delta^2 + |V|^2}$. 

As can be easily shown, the Floquet states $|\Phi_{\varepsilon_m} \rangle$ are eigenfunctions of the both operators $ H_F$ and ${\mathbf S}\cdot \hat {\mathbf R} $,
\begin{align}\label{Eq1e}
 H_F|\Phi_{\varepsilon_m} \rangle =  \varepsilon_m |\Phi_{\varepsilon_m} \rangle, \\
({\mathbf S}\cdot \hat {\mathbf R}) |\Phi_{\varepsilon_m} \rangle = m|\Phi_{\varepsilon_m} \rangle,
\end{align}
where $\varepsilon_m = -j\omega + m\Omega$ $(m=-j,-j+1,\dots,j -1,j) $.  These relations, show that the monopole spin is an observable that can be measured simultaneously with the quasienergy.

The DP is defined by condition $\Omega =0$. As one can see, at the DP  one has a $m$-fold degeneracy. Applying (\ref{Eq28}), we obtain the geometric phase related with the quasienergy $\varepsilon_m$ as $\gamma_{m} =2\pi  (j  + m\cos\theta)$. In particular, for $m = \pm j$, we obtain $\gamma_{\pm j} =2\pi j (1  \pm\cos\theta)$. This is exactly the expression for the Aharonov-Anandan geometric phase for spin-$j$ system \cite{AA}.  As known, in the adiabatic approximation it becomes the Berry phase.

\section{Further generalization}

Here we consider more general case when $n$-levels ($n > 2$) become degenerate at some point. In this situation the emerging monopole has more complicated structure.  Generally, being $SU(n)$ monopole, it has matrix-valued ``charge", and thus can not be characterized by only one quantum number \cite{CSB}.

\subsection{Non-Abelian $SU(n)$ monopoles}

An outline of the $SU(n)$ monopoles theory being provided here is based on the review by S. Coleman \cite{CSB}. Additional details can be found in the aforementioned review, and in the book \cite{SHY}. 

Generalization of the Abelian monopole theory on the $SU(n)$ gauge group is rather straightforward. For instance, the potential with the south-pointing string can be written as 
\begin{align}
A=  Q(1- \cos\theta) d\varphi,
\end{align}
where $Q$ is matrix-valued ``charge" with values in the Lie algebra of the group $SU(n)$, $Q\in {\mathfrak s }{\mathfrak  u}(n)$. The potential with the north-pointing string is defined by\begin{align}
A'=  -Q(1+ \cos\theta) d\varphi.
\end{align}
Both potentials yield the same `magnetic' field,
\begin{align}
\mathbf B = Q\frac{\mathbf R}{R^3}.
\end{align}

The transformation $A \rightarrow  A'$  is given by gauge transformation 
\begin{align}
g=\exp(i2Q\varphi).
\end{align}
The requirement $g$ to be a single-valued function leads to the quantization condition for non-Abelian monopole:
\begin{align}\label{Q3}
\exp(i4\pi Q) = {1\hspace{-.15cm}1}.
\end{align}
Since  $Q\in {\mathfrak s }{\mathfrak  u}(n)$, it must be traceless Hermitian $n\times n$-matrix, 
which can be diagonalized, so that one can obtain
\begin{align}\label{Q1}
Q= -\frac{1}{2}{\rm diag}\{q_1,q_2,\dots,q_n\},
\end{align}
where $\sum^n_{i=1} q_i=0$. The quantization condition implies that all $q_i$'s are integers.

If the gauge group is not $SU(n)$ but $SU(n)/{\mathbb Z}_n$, then the quantization condition
is modified as follows \cite{CSID}
\begin{align}\label{Q3a}
\exp(i4\pi Q_{\rm fund}) \in {\mathbb Z}_n,
\end{align}
where the matrix $Q_{\rm fund}$ denotes the generator  in the $n$-dimensional fundamental representation of the group $SU(n)$. From here it follows
\begin{align}\label{Q4}
Q= -\frac{1}{2}{\rm diag}\{q_1,q_2,\dots,q_n\}.
\end{align}
However, now one has
\begin{align}
q_m = \frac{p}{n} +\rm integer,
\end{align}
where $p$ is an integer, and as before $\sum^n_{m=1} q_m=0$.

\subsection{Trapped $\Lambda$-type atom and $SU(3)$ monopoles}

One possible way to implement a non-Abelian monopole is by employing atom-light interaction of atoms with degenerate internal degrees of freedom coupled to spatially varying laser fields \cite{OKBM,RJF,LHY,LXLX}. 

\begin{figure}[tbh]
\begin{center}
\scalebox{0.4}{\includegraphics{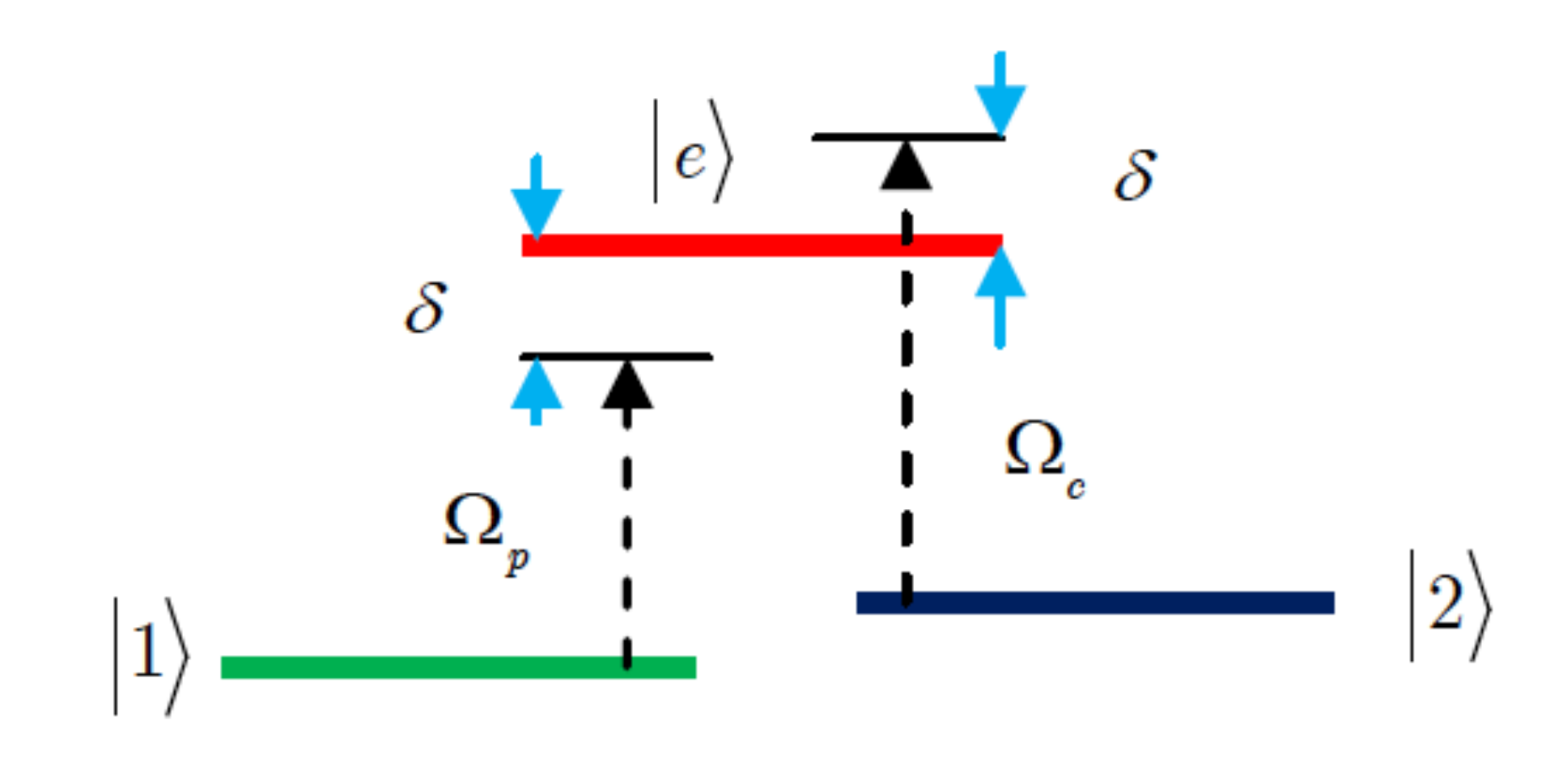}}
\end{center}
\caption{Three level atom interecting with laser beams coupling the  states $|1\rangle$ and $|2\rangle$  to the state $|e\rangle$ .}
\label{L1}
\end{figure}

To illustrate these ideas, we consider a system of trapped $\Lambda$-type atoms interacting with two laser beams, as schematically shown in Fig. \ref{L1} \cite{ZLS}. (For details, we refer the reader to original papers and review \cite{DGJ}.). 
The quantum mechanical system is governed by the Hamiltonian that in the rotating wave approximation can be written as 
\begin{eqnarray}\label{lambdaH}
H=\frac{\mathbf{p}^2}{2m} + V(\mathbf{r}) + H_f(\mathbf r),
\end{eqnarray}
where $\mathbf{p}$ is atom's momentum and $V(\mathbf{r})$ is the trapping potential. The coupling Hamiltonian $H_f$ takes the form,
\begin{align}
H_f = -\frac{\delta}{2}|\mathrm{1}\rangle\langle\mathrm{1}|
+ \frac{\delta}{2}|\mathrm{2}\rangle\langle\mathrm{2}| 
+\frac{\Omega_p}{2}|\mathrm{e}\rangle\langle\mathrm{1}|+\frac{\Omega_c}{2}|\mathrm{e}\rangle\langle\mathrm{2    }|
+ \mbox{h. c.}
\end{align}
Here $\Omega_p$ and $\Omega_c$ are the complex, space-dependent Rabi frequencies for the probe and control laser beams, respectively, and $\delta$ is the one-photon detunning. 

\begin{figure}[tbh]
\scalebox{0.5}{\includegraphics{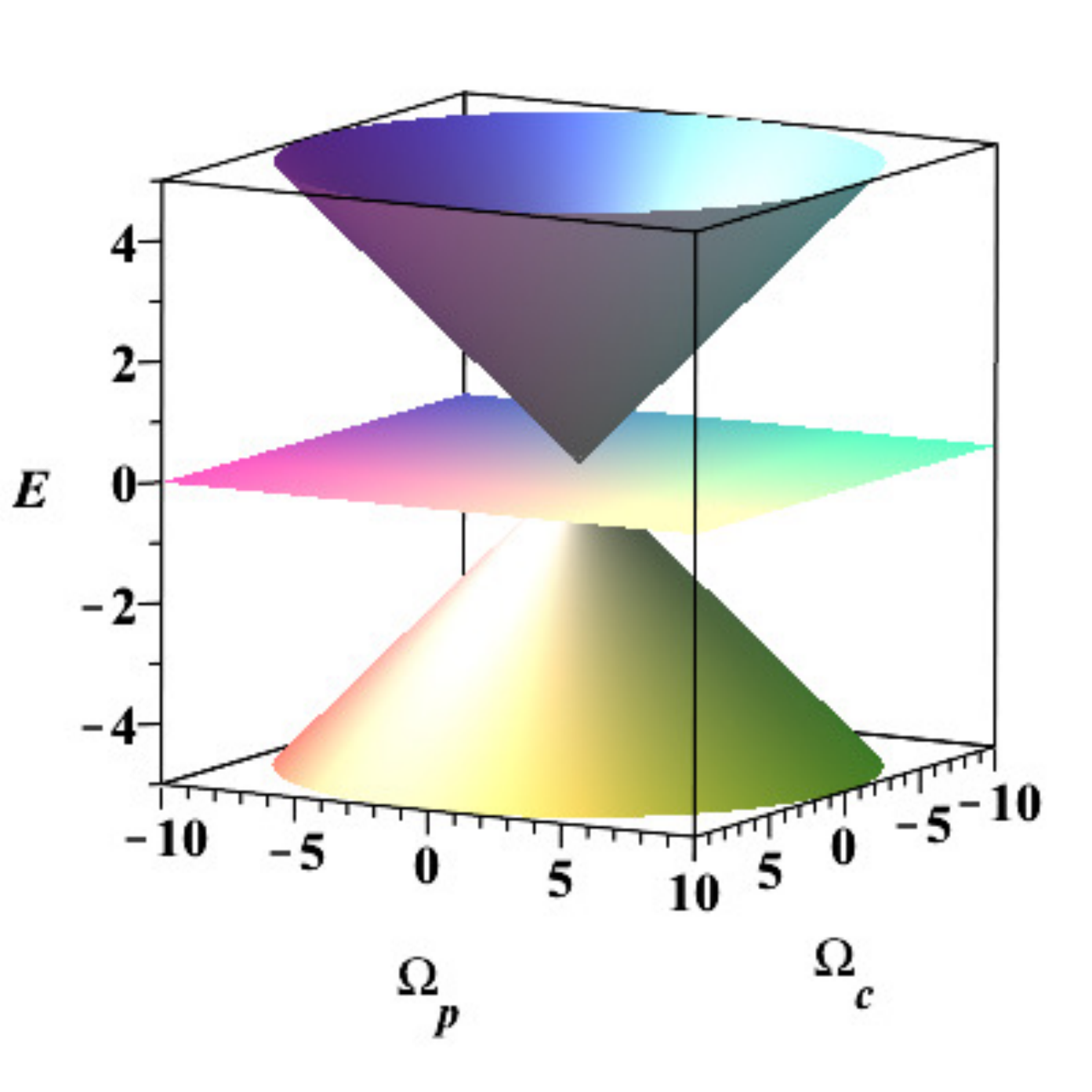}}
\caption{Energy surfaces crossing at the DP. 
\label{L1d}}
\end{figure}

Suppose that that the two-foton excitation is resonant $(\delta =0)$.
In such conditions, the diagonalization of $H_f $ yields the following eigenenergies: $E_0= 0$ and
$E_\pm = \pm R/2$, where $R = \sqrt{|\Omega_p|^2 + |\Omega_c|^2}$.  We consider the complex Rabi frequencies $\Omega_p$ and $\Omega_c$ as control parameters living in ${\mathbb R}^4$, and set $\Omega_p = X + iY$ and $\Omega_c = Z + iU$, where $(X,Y,Z,U )\in {\mathbb R}^4$. As one can see, the surfaces of the constant energies, $|E_{\pm}| = \rm const$, foliate ${\mathbb R}^4$ by $S^3$. 

The eigenstate with the zero energy is called {\em dark} sate,
\begin{align}
|D\rangle = \frac{1}{R}(\Omega_p|\mathrm{2}\rangle -\Omega_c|\mathrm{1}\rangle).
\end{align}
The remaining states with energies $E_\pm$ read 
\begin{align}\label{B1a}
|\pm\rangle =(|B\rangle \pm |e\rangle)/\sqrt{2},
\end{align}
 where $|B\rangle$ is the {\em bright} state,
\begin{align}
|B\rangle = \frac{1}{R}(\Omega^\ast_p|\mathrm{1}\rangle +\Omega^\ast_c|\mathrm{2}\rangle)
\end{align}

If one chooses the Rabi frequencies as
\begin{eqnarray}
\Omega_c &=& |\Omega_c|\mbox{e}^{i\varphi_c},\\
\Omega_p &=& |\Omega_p|\mbox{e}^{i\varphi_p},
\end{eqnarray}
then in the $|\mathrm{1}\rangle,|\mathrm{2}\rangle,|\mathrm{e}\rangle$ basis the dark and the bright states can be written as
\begin{align}
|D\rangle =  \left(\begin{array}{c}
-e^{i\varphi_c}\cos\frac{\theta}{2}\\
e^{i\varphi_p}\sin\frac{\theta}{2} \\
0
\end{array}\right),
 \label{D1}\\
 |B\rangle = \left(\begin{array}{c}
                  e^{-i\varphi_p}\sin\frac{\theta}{2}\\
                  e^{-i\varphi_c}\cos\frac{\theta}{2}\\
                  0
                  \end{array}\right),
 \label{D2}
\end{align}
where 
\begin{align}
\tan \frac{\theta}{2} =\frac{|\Omega_p|}{|\Omega_c|}.
\end{align}

Making use of these results for calculation of the induced gauge field we obtain
\begin{align} \label{A1a}
A_0 = \langle D|d|D\rangle =&-\frac{1}{2}( d\psi -\cos \theta d\varphi), \\
A_\pm =\langle \pm|d|\pm\rangle=& \frac{1}{4}(d\psi- \cos \theta d\varphi ) .
\label{A2a}
\end{align}
Here we set  $\varphi =\varphi_p -\varphi_c$ and $\psi = \varphi_p +\varphi_c$. One can gather all terms and recast  Eqs. (\ref{A1a}), (\ref{A2a}) as,
\begin{align}
A = Q_0(d\psi- \cos \theta d\varphi ),
\label{EqC}
\end{align}
where
\begin{align} \label{Q2}
Q_0=\frac{1}{2}{\rm diag}\bigg\{-1, \frac{1}{2}, \frac{1}{2}\bigg\}.
\end{align}

Eq. \eqref{EqC} defines a natural or canonical connection on $S^3$. The corresponding  curvature $F$, yielding a monopole field, reads
\begin{align}
F = dA = Q_0\sin\theta \,d\theta \wedge d\varphi.
\end{align}

{\em Remark.} Geometric phases for two and three-level quantum systems governed by the Hamiltonian belonging to the Lie group ${\mathfrak s }{\mathfrak  u}(2)$ or ${\mathfrak s }{\mathfrak  u}(3)$, respectively, are considered in details in  Refs. \cite{KGMS,KAMN,BAY1}.

Topological structures of adiabatic phase for multi-level quantum systems \cite{LZZX}.

\subsubsection{Laser beams with orbital angular momentum}

Here we consider the laser beams with the orbital angular momentum $l_p$ and $l_c$, so that $\varphi_p = l_p \varphi$ and $\varphi_c = l_c \varphi$ \cite{DGJ}. This implies that the Rabi frequencies can be written as,  $\Omega_c =|\Omega_c|\mbox{e}^{il_c\varphi}$ and
$\Omega_p =|\Omega_p|\mbox{e}^{i l_p\varphi}$. Using these results in (\ref{A1a}) and (\ref{A2a}), we find
\begin{align} \label{A1}
A_0 = &-\frac{l}{2}(1-\cos \theta) d\varphi - l_c d\varphi , \\
A_\pm =& \frac{l}{4}(1- \cos \theta d\varphi ) + \frac{l_c}{2}d\varphi,
\label{A2}
\end{align}
where $l= l_p -l_c$ is the relative winding number of beams. From here we obtain
\begin{align}
F = Q_l\sin\theta \,d\theta \wedge d\varphi, 
\end{align}
where
\begin{align} \label{Q2a}
Q_l=\frac{1}{2}{\rm diag}\bigg\{-l, \frac{l}{2}, \frac{l}{2}\bigg\}.
\end{align}
Below we will show that the charge $Q_l$ belongs to the Lie algebra of the group $SU(3)$, $Q_l\in {\mathfrak s }{\mathfrak  u}(3)$.

We look at the Rabi frequencies as the control parameters, taking them as $|\Omega_c|$, $|\Omega_p|$ and  $\varphi$. Thus  the parameter space is three-dimensional manifold ${\mathfrak M}\in {\mathbb R}^4$. We define a mapping ${\mathfrak M} \mapsto {\mathbb R}^3$ as follows:
\begin{eqnarray}
\Omega_c &=&R\cos \frac{\theta}{2}\,\mbox{e}^{il_c\varphi}, \\
\Omega_p &=& R\sin \frac{\theta}{2}\,\mbox{e}^{i l_p\varphi},
\end{eqnarray}
where $(r,\theta,\varphi)$ are the spherical coordinates in ${\mathbb R}^3$.
The strength of the corresponding monopole field can be written as
\begin{equation}\label{Eq4kg}
{\mathbf B }= Q_l\frac{\mathbf R}{R^3} .
\end{equation}

 An arbitrary monopole charge for $SU(3)$ monopole can be written as follows \cite{SHY}:
\begin{align}
Q=\bigg(n_1 -\frac{n_2}{2} \bigg)\Gamma_1 + \frac{\sqrt{3}}{2} n_2 \Gamma_2,
\label{Q}
\end{align}
where $\Gamma_1$ and $ \Gamma_2$ are generators of  Cartan subalgebra of $SU(3)$,
\begin{align}
\Gamma_1 = \frac{1}{2}
\left(\begin{matrix}
  1& 0&0 \\
  0&-1 &0 \\
  0&0 &0
\end{matrix}
\right), \quad
\Gamma_2 = \frac{1}{2\sqrt{3}}
\left(\begin{matrix}
  1& 0&0 \\
  0&1 &0 \\
  0&0 &-2
\end{matrix}
\right)
\label{G1}
\end{align}
Using (\ref{G1}) in (\ref{Q}), we get
\begin{align}\label{Q1a}
Q=\frac{1}{2}{\rm diag}\big\{n_1, n_2-n_1, -n_2\big\},
\end{align}
where $n_1,n_2$ are integers. Choosing $n_1 = -2l$ and $n_2 =-l$ we obtain
\begin{align}\label{Q3b}
Q=\frac{l}{2}{\rm diag}\big\{-2, 1, 1\big\}.
\end{align}

The spin approach is recovered employing the unitary transformation
\begin{align}
U(\theta,\varphi)=\left(
\begin{array}{ccc}
\cos\frac{\theta}{2} e ^{-i\varphi/2}& \sin\frac{\theta}{2}e ^{i\varphi/2}&0\\ 
-\sin\frac{\theta}{2}e ^{-i\varphi/2}& \cos\frac{\theta}{2}e ^{i\varphi/2}& 0 \\
0&0&1 
\end{array} 
\right ).
\end{align}
Under this transformation, the strength of the monopole field, \eqref{Eq4k}, becomes
\begin{equation}\label{Eq4ka}
{\mathbf B } = - l ({\mathbf S}\cdot \hat {\mathbf R})\frac{\mathbf R}{R^3}  - \frac{l\sqrt{3}}{2} \Gamma_2\frac{\mathbf R}{R^3} .
\end{equation}

For general $SU(3)$ monopole, we obtain
\begin{equation}\label{Eq4kc}
{\mathbf B } = \bigg(n_1 -\frac{n_2}{2} \bigg) ({\mathbf S}\cdot \hat {\mathbf 
R})\frac{\mathbf R}{R^3}  + \frac{n_2\sqrt{3}}{2} \Gamma_2\frac{\mathbf R}{R^3} ,
\end{equation}
where the first term defines the monopole spin, $\mathbf S$, and the second one, $Q_h 
={n_2\sqrt{3}}/{2} \Gamma_2$, describes its hypercharge.

To classify obtained solution one should look for stability group $H \in SU(3)$ of $Q_0$ and determine $\pi_2(SU(3)/H)$ \cite{PJ}. The generators of $H$ should commute with $Q_0$. There are generators of $SU(2)$, that mix the two degenerated values of $Q_0$ and $U(1)$ generator $Q$ given by Eq. (\ref{Q3}). As was shown in \cite{PJ}, 
\begin{align}
H \simeq [SU(2)\times U(1)]/\mathbb{ Z}_2 \simeq U(2).
\end{align}
Next, using the identity $\pi_2(SU(3)/H)=\pi_1(H)$ and the fact that the fundamental group of $U(n)$ is $\pi_1(U(n)) =\mathbb Z$, one can obtain \cite{SHY}
\begin{align}
\pi_2(SU(3)/H)= \mathbb Z.
\end{align}
Thus, in general case the topological charge of the monopole is 
\begin{align}\label{Q1g}
Q_0=\frac{1}{4}{\rm diag}\big\{n_1, n_2-n_1, -n_2\big\},
\end{align}
where $n_1,n_2$ are integers. The quantization condition (\ref{Q3}) takes the form:
\begin{align}\label{Q4a}
\exp(i4\pi Q_0) = {\rm diag}\big\{1,-1,-1\big\}
\end{align}
Comparing this expression with (\ref{Q2}), we find that $Q_0 = Q/2$.

\section*{Conclusion}

In this paper, we studied the artificial magnetic monopoles associated with energy level crossing in quantum systems.  In the simplest case of double degeneracy,  the monopole is located at the degenerate point. Then, in the standard approach, one has two monopoles with similar structures of the magnetic field but with opposite charges. In the more complicated case of $n$-level energy crossing there appear $n$ monopoles related to the corresponding eigenstates. Each monopole has its own charge, and the total charge of all monopoles equals zero \cite{B0,BD,Chj}.

We show that in general case of $n$-level energy crossing the quantum system can be successfully described by a {\em single} non-Abelian artificial `magnetic' monopole with hidden symmetries and isospin degrees of freedom. In initial representation of the quantum system these hidden symmetries reveal themselves as a mixture of spin and other degrees of freedom. 

In particular, for two-level system this monopole behaves as a quasi-particle with a spin one half. We demonstrated by different examples how the artificial monopole exposes itself. For a trapped $\Lambda$-type system we found the $SU(3)$ monopole with the hidden symmetry $U(2) \simeq [SU(2)\times U(1)]/\mathbb{ Z}_2$. In this case, the monopole charge  is fractional and represents the mixture of spin and Abelian charge degrees of freedom. These new features of artificial monopoles offer the possibility of the experimental test and can be useful in many physical applications. 

\begin{acknowledgements}
AIN acknowledges the support from the CONACyT.
The work by G.P.B. was done at Los Alamos National Laboratory managed by Triad National Security, LLC, for the National Nuclear Security Administration of the U.S. Department of Energy under Contract No. 89233218CNA000001.
 \end{acknowledgements}


\begin{thebibliography}{10}
\providecommand{\url}[1]{{#1}}
\providecommand{\urlprefix}{URL }
\expandafter\ifx\csname urlstyle\endcsname\relax
  \providecommand{\doi}[1]{DOI \discretionary{}{}{}#1}\else
  \providecommand{\doi}{DOI \discretionary{}{}{}\begingroup
  \urlstyle{rm}\Url}\fi

\bibitem{BM}
E.~Barouch, B.M. McCoy, Phys. Rev. A \textbf{3}, 786 (1971)

\bibitem{Br}
R.~Bhandari, Phys. Rev. Lett. \textbf{89}, 268901 (2002)

\bibitem{FNT}
{Z. Fang {\it et al}}, Science \textbf{302}, 92 (2003)

\bibitem{SR}
C.M. Savage, J.~Ruostekoski, Phys. Rev. A \textbf{68}, 043604 (2003)

\bibitem{Hal}
F.D.M. Haldane, Phys. Rev. Lett. \textbf{93}, 206602 (2004)

\bibitem{FP}
{J. Frenkel and S. H. Pereira}, Phys. Rev. D \textbf{69}, 127702 (2004)

\bibitem{ZLS}
P.~Zhang, Y.~Li, C.P. Sun, Eur. Phys. J. D \textbf{36}, 229  (2005)

\bibitem{MSN}
S.~Morita, H.~Nishimori, J. Math. Phys. \textbf{49}, 125210 (2008)

\bibitem{PVMM}
V.~Pietil\"a, M.~M\"ott\"onen, Phys. Rev. Lett. \textbf{102}, 080403 (2009)

\bibitem{DGJ}
J.~Dalibard, F.~Gerbier, G.~Juzeli\ifmmode~\bar{u}\else \={u}\fi{}nas,
  P.~\"Ohberg, Rev. Mod. Phys. \textbf{83}, 1523 (2011)

\bibitem{RPN}
P.~Ra\ifmmode~\check{c}\else \v{c}\fi{}kauskas, V.~Novi\ifmmode~\check{c}\else
  \v{c}\fi{}enko, H.~Pu, G.~Juzeli\ifmmode~\bar{u}\else \={u}\fi{}nas, Phys.
  Rev. A \textbf{100}, 063616 (2019)

\bibitem{NVJ}
V.~Novi\ifmmode~\check{c}\else \v{c}\fi{}enko, G.~Juzeli\ifmmode~\bar{u}\else
  \={u}\fi{}nas, Phys. Rev. A \textbf{100}, 012127 (2019)

\bibitem{RVWC}
G.~Valent\'{\i}-Rojas, N.~Westerberg, P.~\"Ohberg, Phys. Rev. Research
  \textbf{2}, 033453 (2020)

\bibitem{CSPA}
S.~Cusumano, A.~De~Pasquale, V.~Giovannetti, Phys. Rev. Lett. \textbf{124},
  190401 (2020)

\bibitem{RRKMH}
M.W. Ray, E.~Ruokokoski, M.~Kandel, S.~Mottonen, D.S. Hall, Nature
  \textbf{505}, 657–660 (2014)

\bibitem{B0}
M.V. Berry, Proc. R. Soc. A \textbf{392}, 45  (1984)

\bibitem{B1}
M.V. Berry, in \emph{Anomalies, Phases, Defects}, ed. by U.M. Bregola,
  G.~Marino, G.~Morandi (Bibliopolis, Naples, 1990), p. 125

\bibitem{Arn}
V.I. Arnold, \emph{Geometric Methods in the Theory of Ordinary Differential
  Equations} (Springer, New York, 1983)

\bibitem{BW}
M.V. Berry, M.~Wilkinson, Proc. Roy. Soc. A \textbf{392}, 15  (1984)

\bibitem{BD}
M.V. Berry, M.R. Dennis, Proc. Roy. Soc. A \textbf{459}, 1261  (2003)

\bibitem{AA}
Y.~Aharonov, J.~Anandan, Phys. Rev. Lett. \textbf{58}(16), 1593 (1987)

\bibitem{WFZA}
F.~Wilczek, A.~Zee, Phys. Rev. Lett. \textbf{52}, 2111 (1984)

\bibitem{MCA}
C.A. Mead, Phys. Rev. Lett. \textbf{59}, 161 (1987)

\bibitem{MCA1}
C.A. Mead, Rev. Mod. Phys. \textbf{64}, 51 (1992)

\bibitem{WFSA}
A.~Shapere, F.~Wilczek (eds.), \emph{Geometric Phases in Physics} (World Sci.,
  Singapore, 1989)

\bibitem{Dir}
P.A.M. Dirac, Proc. Roy. Soc. Lond. A \textbf{133}, 60  (1931)

\bibitem{Sw_1}
J.~Schwinger, Phys. Rev. \textbf{144}, 1087 (1966)

\bibitem{NF}
A.I. Nesterov, F.~{A}ceves de~la Cruz, Phys. Lett. A \textbf{302}, 253  (2002)

\bibitem{N1a}
A.I. Nesterov, Phys. Lett. A \textbf{328}, 110  (2004)

\bibitem{Wu2}
T.T. Wu, C.N. Yang, Nucl. Phys. B \textbf{107}, 365  (1976)

\bibitem{Jac}
R.~Jackiw, Phys. Rev. Lett. \textbf{54}, 159  (1985)

\bibitem{Jac1}
R.~Jackiw, Phys. Lett. B \textbf{154}, 303  (1985)

\bibitem{NG}
G.~Nadeau, Am. J. Phys. \textbf{28}, 566 (1960)

\bibitem{LRP}
I.R. I.~Richard~Lapidus, J.L. Pietenpol, Am. J. Phys. \textbf{28}, 17 (1960)

\bibitem{LBNZ}
D.~Lynden-Bell, M.~Nouri-Zonoz, Rev. Mod. Phys. \textbf{70}, 427 (1998)

\bibitem{SHY}
Y.~Shnir, \emph{Magnetic Monopoles} (Springer, New York, 2005)

\bibitem{WHA}
H.A. Wilson, Phys. Rev. \textbf{75}, 309 (1949)

\bibitem{Gol1}
A.S. Goldhaber, Phys. Rev. \textbf{140}, B1407  (1965)

\bibitem{Gol2}
A.S. Goldhaber, Phys. Rev. Lett. \textbf{36}, 1122  (1976)

\bibitem{SMN}
M.N. Saha, Indian J. Phys. \textbf{10}, 145 (1936)

\bibitem{SMN1}
M.N. Saha, Phys. Rev. \textbf{75}, 1968 (1949)

\bibitem{WF}
F.~Wilczek, Phys. Rev. Lett. \textbf{48}, 1144 (1982)

\bibitem{WF1}
F.~Wilczek, Phys. Rev. Lett. \textbf{48}, 1146 (1982)

\bibitem{JRRC}
R.~Jackiw, C.~Rebbi, Phys. Rev. Lett. \textbf{36}, 1116 (1976)

\bibitem{HPHG}
P.~Hasenfratz, G.~'t~Hooft, Phys. Rev. Lett. \textbf{36}, 1119 (1976)

\bibitem{SJ1}
J.~Schwinger, Science \textbf{165}, 757 (1969)

\bibitem{MKD}
K.A. Milton, L.L. DeRaad, J. Math. Phys. \textbf{19} (1978)

\bibitem{MAK}
K.A. Milton, Rep. Prog. Phys. \textbf{69}, 1637 (2006)

\bibitem{PD}
D.N. Page, Phys. Rev. A \textbf{36}, 3479 (1987)

\bibitem{SB}
B.~Simon, Phys. Rev. Lett. \textbf{51}, 2167 (1983)

\bibitem{Chj}
D.C.A. Jamiolkowski, \emph{{ Geometric Phases in Classical and Quantum
  Mechanics}} (Birkh{\"a}user, New York, 2004)

\bibitem{AS}
J.~Anandan, L.~Stodolsky, Phys. Rev. D \textbf{35}, 2597 (1987)

\bibitem{MMSY}
M.~Maamache, Y.~Saadi, Phys. Rev. Lett. \textbf{101}, 150407 (2008)

\bibitem{MS}
S.~Massar, Phys. Rev. A \textbf{54}, 4770 (1996)

\bibitem{MS1}
N.~Mukunda, R.~Simon, Ann. Phys. \textbf{228}, 205 (1993)

\bibitem{MS2}
N.~Mukunda, R.~Simon, Ann. Phys. \textbf{228}, 269  (1993)

\bibitem{GW}
J.C. Garrison, E.M. Wright, Phys. Lett. A \textbf{128}, 177  (1988)

\bibitem{B2}
M.V. Berry, Ann. N. Y. Acad. Sci. \textbf{755}, 303  (1995)

\bibitem{KMS}
O.N. Kirillov, A.A. Mailybaev, A.P. Seyranian, J. Phys. A \textbf{38}, 5531
  (2005)

\bibitem{HPTHG}
P.~Hasenfratz, G.~'t~Hooft, Phys. Rev. Lett. \textbf{36}, 1119 (1976)

\bibitem{MSW}
J.~Moody, A.~Shapere, F.~Wilczek, Phys. Rev. Lett. \textbf{56}, 893 (1986)

\bibitem{RJRC}
R.~Jackiw, C.~Rebbi, Phys. Rev. Lett. \textbf{36}, 1116 (1976)

\bibitem{LHZ}
H.Z. Li, Phys. Rev. Lett. \textbf{58}, 539 (1987)

\bibitem{LHZ1}
H.Z. Li, Phys. Rev. D \textbf{35}, 2615 (1987)

\bibitem{JR1}
R.~Jackiw, Phys. Rev. Lett. \textbf{56}, 2779 (1986)

\bibitem{Hooft}
{G.'t Hooft}, Nuclear Physics B \textbf{79}, 276  (1974)

\bibitem{PAM}
A.~Polyakov, Jetp Letters \textbf{20}, 194 (1974)

\bibitem{AJFP}
J.~Arafune, P.G.O. Freund, C.J. Goebel, J. Math. Phys. \textbf{16}, 433 (1975)

\bibitem{RVI}
V.I. Ritus, Zh. Eksp. Ter. Fiz. \textbf{51}, 1544 (1966)

\bibitem{BPXH}
P.~Brusheim, H.Q. Xu, Phys. Rev. B \textbf{79}, 205323 (2009)

\bibitem{SHI}
H.~Sambe, Phys. Rev. A \textbf{7}, 2203 (1973)

\bibitem{HGM}
J.~Hausinger, M.~Grifoni, Phys. Rev. A \textbf{81}, 022117 (2010)

\bibitem{MDJ}
D.J. Moore, J. Phys. A \textbf{23}, L665 (1990)

\bibitem{MDJ1}
D.J. Moore, J. Phys. A \textbf{23}, 5523 (1990)

\bibitem{MDJ2}
D.J. Moore, Phys. Rep. \textbf{210}, 1  (1991)

\bibitem{GMPH}
M.~Grifoni, P.~Hanggi, Phys. Rep. \textbf{304}, 229  (1998)

\bibitem{AK}
K.~Aizu, J. Math. Phys. \textbf{4}, 125210 (1963)

\bibitem{BWB}
W.B. Brown, Proc. Cambridge. Phil. Soc. \textbf{54}, 251 (1958)

\bibitem{SJH}
J.H. Shirley, Phys. Rev. \textbf{138}, B979 (1965)

\bibitem{LHCH}
E.~Layton, Y.~Huang, S.I. Chu, Phys. Rev. A \textbf{41}, 42 (1990)

\bibitem{CSB}
S.~Coleman, \emph{Aspects of Symmetry: Selected Erice Lectures} (Cambridge
  University Press, Cambridge, 1995)

\bibitem{CSID}
S.~Coleman, in \emph{Les Houches 1981, Proceedings, Gauge Theories in High
  Energy Physics, Part 1} (1983)

\bibitem{OKBM}
K.~Osterloh, M.~Baig, L.~Santos, P.~Zoller, M.~Lewenstein, Phys. Rev. Lett.
  \textbf{95}, 010403 (2005)

\bibitem{RJF}
J.~Ruseckas, G.~Juzeli\ifmmode~\bar{u}\else \={u}\fi{}nas, P.~\"Ohberg,
  M.~Fleischhauer, Phys. Rev. Lett. \textbf{95}, 010404 (2005)

\bibitem{LHY}
L.H. Lu, Y.Q. Li, Phys. Rev. A \textbf{76}, 023410 (2007)

\bibitem{LXLX}
X.J. Liu, X.~Liu, L.C. Kwek, C.H. Oh, Phys. Rev. Lett. \textbf{98}, 026602
  (2007)

\bibitem{KGMS}
G.~Khanna, S.~Mukhopadhyay, R.~Simon, N.~Mukunda, Annals of Physics
  \textbf{253}, 55 (1997)

\bibitem{KAMN}
Arvind, K.S. Mallesh, N.~Mukunda, Journal of Physics A: Mathematical and
  General \textbf{30}(7), 2417 (1997)

\bibitem{BAY1}
Y.~Ben-Aryeh, Optics and Spectroscopy \textbf{94}(5), 724 (2003)

\bibitem{LZZX}
Z.~Liu, X.~Zhou, X.~Liu, J.~Chen, Journal of Physics A \textbf{40}, 1661 (2007)

\bibitem{PJ}
{J. Preskill}, Ann. Rev. of Nucl. and Part. Sci. \textbf{34}, 461  (1984)

\end{thebibliography}
\end{document}